\documentclass[12pt,twocolumn,tighten]{aastex63}
\usepackage{amsmath,amstext,amssymb}
\usepackage[T1]{fontenc}
\usepackage{apjfonts}
\usepackage[figure,figure*]{hypcap}
\usepackage{graphics,graphicx}
\usepackage{hyperref}
\usepackage{natbib}
\usepackage[caption=false]{subfig} 
\usepackage{enumitem} 
\usepackage{epigraph}

\newcommand{\npms}{1{,}097} 
\newcommand{\nchtwo}{37} 
\newcommand{\nrsgfive}{173} 

\newcommand{\kms}{\,km\,s$^{-1}$}
\newcommand{\mkms}{{\rm \,km\,s^{-1}}}  

\newcommand{\bpmrpo}{(G_{\rm BP}-G_{\rm RP})_0}

\received{2022 May 2}
\revised{2022 Sep 14}
\accepted{2022 Sep 17}
\submitjournal{Astronomical Journal}
\shorttitle{Kepler and the Behemoth}

\begin{document}

\title{
  Kepler and the Behemoth: Three Mini-Neptunes in a 40 Million Year Old Association
}

\correspondingauthor{L.\,G.\,Bouma}
\email{luke@astro.caltech.edu}

\author[0000-0002-0514-5538]{L. G. Bouma}
\altaffiliation{51 Pegasi b Fellow}
\affiliation{Cahill Center for Astrophysics, California Institute of Technology, Pasadena, CA 91125, USA}

\author[0000-0002-6549-9792]{R.~Kerr} 
\affiliation{Department of Astronomy, The University of Texas at Austin, Austin, TX 78712, USA}
%
\author[0000-0002-2792-134X]{J. L. Curtis} 
\affiliation{Department of Astronomy, Columbia University, 550 West 120th Street, New York, NY 10027, USA}
%
\author[0000-0002-0531-1073]{H. Isaacson} 
\affiliation{Astronomy Department, University of California, Berkeley, CA 94720, USA}
%
\author{L. A. Hillenbrand} 
\affiliation{Cahill Center for Astrophysics, California Institute of Technology, Pasadena, CA 91125, USA}
%
\author[0000-0001-8638-0320]{A. W. Howard} 
\affiliation{Cahill Center for Astrophysics, California Institute of Technology, Pasadena, CA 91125, USA}
%
\author[0000-0001-9811-568X]{A.~L.~Kraus} 
\affiliation{Department of Astronomy, The University of Texas at Austin, Austin, TX 78712, USA}
%
\author[0000-0001-6637-5401]{A. Bieryla} 
\affiliation{Center for Astrophysics \textbar \ Harvard \& Smithsonian, 60 Garden St, Cambridge, MA 02138, USA}
%
\author[0000-0001-9911-7388]{D. W.~Latham} 
\affiliation{Center for Astrophysics \textbar \ Harvard \& Smithsonian, 60 Garden St, Cambridge, MA 02138, USA}
%
\author[0000-0003-0967-2893]{E. A.~Petigura} 
\affiliation{Department of Physics \& Astronomy, University of California Los Angeles, Los Angeles, CA 90095, USA}
%
\author[0000-0001-8832-4488]{D. Huber} 
\affiliation{Institute for Astronomy, University of Hawai`i, 2680 Woodlawn Drive, Honolulu, HI 96822, USA}

\begin{abstract}
  Stellar positions and velocities from Gaia are yielding a new view
  of open cluster dispersal.  
  Here we present an analysis of a group
  of stars spanning Cepheus ($l=100^\circ$) to Hercules ($l=40^\circ$),
  hereafter the Cep-Her complex.  The group includes four Kepler
  Objects of Interest:
  Kepler-1643 b ($R_{\rm p} = 2.32 \pm 0.13\,R_\oplus$, $P = 5.3\ {\rm days}$),
  KOI-7368 b ($R_{\rm p} = 2.22 \pm 0.12\,R_\oplus$, $P = 6.8\ {\rm days}$), 
  KOI-7913 Ab ($R_{\rm p} = 2.34 \pm 0.18\,R_\oplus$, $P = 24.2\ {\rm days}$), and
  Kepler-1627 Ab ($R_{\rm p} = 3.85 \pm 0.11\,R_\oplus$, $P = 7.2\ {\rm days}$).
  The latter Neptune-sized planet is in part of the Cep-Her complex
  called the $\delta$\ Lyr\ cluster \citep{bouma_kep1627_2022}.  Here
  we focus on the former three systems, which are in other regions of
  the association.  Based on kinematic evidence from Gaia, stellar
  rotation periods from TESS, and spectroscopy, these three objects
  are also $\approx$40 million years (Myr) old.  More specifically, we
  find that Kepler-1643 is $46^{+9}_{-7}$\,Myr old, based on its
  membership in a dense sub-cluster of the complex called RSG-5.
  KOI-7368 and KOI-7913 are $36^{+10}_{-8}$\,Myr old, and are in a
  diffuse region that we call CH-2.  Based on the transit shapes and
  high resolution imaging, all three objects are most likely planets,
  with false positive probabilities of $6\times10^{-9}$,
  $4\times10^{-3}$, and $1\times10^{-4}$ for Kepler-1643, KOI-7368,
  and KOI-7913 respectively.  These planets demonstrate
  that mini-Neptunes with sizes of $\approx$2 Earth radii exist at ages
  of 40 million years.
\end{abstract}



\section{Introduction}

The discovery and characterization of planets younger than a billion
years is a major frontier in current exoplanet research.  The reason
is that the properties of young planets provide benchmarks for studies
of planetary evolution.  For instance, young planets can inform our
understanding of when hot Jupiters arrive on their close-in orbits
\citep{dawson_johnson_2018}, how the sizes of planets with massive
gaseous envelopes evolve \citep{rizzuto_tess_2020}, the timescales for
close-in multiplanet systems to fall out of resonance
\citep{izidoro_breaking_2017,arevalo_stability_2022,goldberg_architectures_2022}, and
whether and how mass-loss explains the radius valley
\citep{lopez_how_2012,Owen_Wu_2013,Fulton_et_al_2017,ginzburg_corepowered_2018,lee_primordial_2021}.

The discovery of a young planet requires two claims to be true: the
planet must exist, and its age must be secured.  Spaced-based
photometry from K2 and TESS has yielded a number of exemplars for
which the planetary evidence comes from transits, and the age is based
on either cluster membership
\citep{Mann_et_al_2017,david_four_2019,newton_tess_2019,bouma_cluster_2020,nardiello_pathosII_2020}
or else on correlates of youth such as stellar rotation, photospheric
lithium content, x-ray activity, and emission line strength
\citep{zhou_2021_tois,hedges_toi-2076_2021}.

In this work, we leverage recent analyses of the Gaia data, which have
greatly expanded our knowledge of stellar groups
\citep[{e.g.},][]{CantatGaudin2018a,KounkelCovey2019,Kerr2021}.
So far, these analyses have mostly
leveraged 3D stellar positions and 2D on-sky tangential velocities.
One important result has been the discovery of diffuse
streams and tidal tails comparable in stellar mass to the previously
known cores of nearby open clusters
\citep{meingast_psceri_2019,Meingast2021,gagne_number_2021}.  Even
though these streams are spread over tens to hundreds of parsecs,
their velocity dispersions can remain coherent at the $\sim$1\kms\
level.  Internal dynamics and projection effects can also drive them
to be much more kinematically diffuse: in the Hyades, stars in the tidal tails are
expected to span up to $\pm 40\mkms$ in velocity relative to the
cluster center \citep{jerabkova_800_2021}.  The stars in such diffuse
regions can be verified to be the same age as the core cluster members
through analyses of color--absolute magnitude diagrams
\citep{KounkelCovey2019}, stellar rotation periods
\citep{curtis_tess_2019,bouma_2021_ngc2516}, and chemical abundances
\citep{hawkins_2020,2020A&A...635L..13A}.  While there are implications for our
understanding of star formation and cluster evolution
\citep{dinnbier_tidal_2020}, a separate consequence is that we now
know the ages of many more stars, including previously known planet
hosts.

The prime Kepler mission \citep{borucki_kepler_2010} found most of the
currently known transiting exoplanets, and it was conducted before
Gaia.  It is therefore sensible to revisit the Kepler field, given our
new constraints on the stellar ages.

Here, we expand on our earlier study of a $38^{+7}_{-6}$ Myr old
Neptune-sized planet in the Kepler field (Kepler-1627~Ab;
\citealt{bouma_kep1627_2022}).  This planet's age was derived based on
its host star's membership in the $\delta$\ Lyr\ cluster.  While our
analysis of the cluster focused on the immediate vicinity of
Kepler-1627 in order to have a reasonable scope,  it became clear that
the $\delta$\ Lyr\ cluster seems to also be part of a much larger
group of similarly aged stars.  This association, which is at an
average distance of 330\,pc from the Sun, spans Cepheus to Hercules
(galactic longitudes, $l$, between 40$^\circ$ and 100$^\circ$), at
galactic latitudes between 0$^\circ$ and 20$^\circ$.  An assessment of
its membership, substructure, and age distribution will be provided as
part of the 1\,kpc expansion of the SPYGLASS project (R. Kerr et al.\
in prep), where it is given the name Cep-Her, after the endpoint
constellations.

Our focus is on the intersection of the Cep-Her complex with the
Kepler field.  Cross-matching the stars thought to be in Cep-Her against known Kepler
Objects of Interest (KOIs; \citealt{thompson_planetary_2018}) yielded
four candidate cluster members: Kepler-1627, Kepler-1643, KOI-7368,
and KOI-7913.  Given our earlier analysis of Kepler-1627, we focus
here on the latter three objects.  After analyzing the relevant
properties of Cep-Her (Section~\ref{sec:cluster}), we derive the
stellar properties (Section~\ref{sec:stars}) and validate the
planetary nature of each system using a combination of the Kepler
photometry and high-resolution imaging (Section~\ref{sec:planets}).
We conclude with a discussion of mini-Neptune size evolution, and
point out possible directions for future work
(Section~\ref{sec:disc_conc}).

\section{The Cep-Her Complex}
\label{sec:cluster}

\begin{figure*}[t]
	\begin{center}
		\leavevmode
		\includegraphics[width=0.99\textwidth]{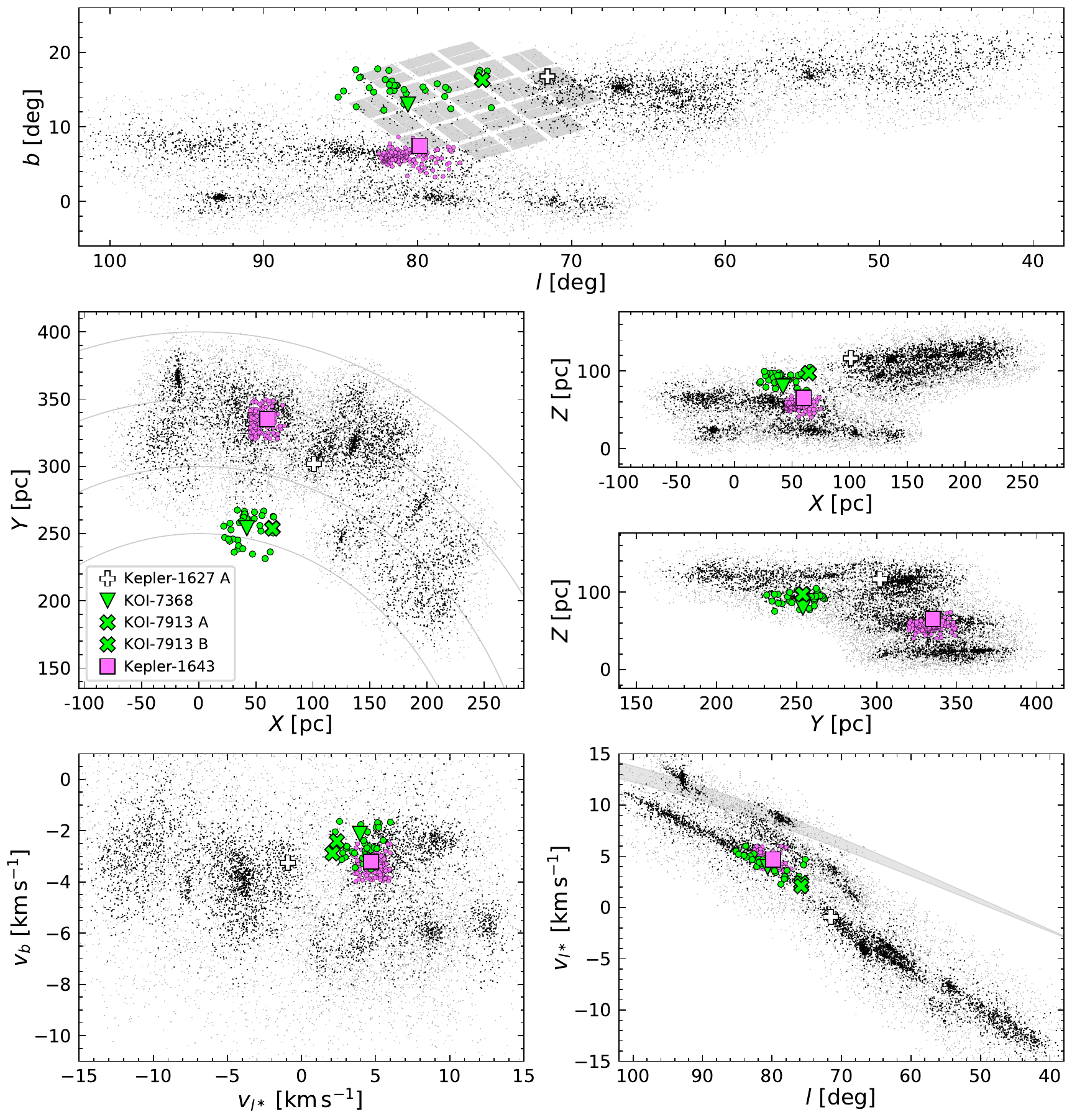}
	\end{center}
	\vspace{-0.6cm}
	\caption{
  {\bf Positions and velocities of candidate members of the Cep-Her
  complex.}
  {\it Top row}: On-sky positions in galactic coordinates.  Black
  points are stars for which group membership is more secure than for
  gray points.  Kepler-1627 is in the outskirts of the $\delta$\ Lyr
  cluster \citep{bouma_kep1627_2022}, which is centered at $(l,b)
  \approx (66^\circ, 12^\circ)$.
  The Kepler footprint is shown in gray.
  {\it Middle row}: Galactic positions.  The Sun is at $(X, Y, Z) =
  (0, 0, 20.8)$\,pc; lines of constant heliocentric distance are
  shown between 250 and 400\,pc, spaced by 50\,pc.
  {\it Bottom row}: Galactic tangential velocities (left) and
  galactic longitudinal velocity versus galactic longitude (right).
  The gray band in the lower-right shows the $\pm$1-$\sigma$
  projection of the Solar velocity with respect to the local standard
  of rest \citep{schonrich_local_2010}.  There is a strong spatial and kinematic overlap between
  Kepler-1643 and RSG-5 (magenta; smaller circles).  The local population
  of candidate young stars around KOI-7368 and KOI-7913 is more
  diffuse -- we call this region ``CH-2'' (lime-green; larger
  circles).
  The selection method for these groups
  is described in Section~\ref{subsec:members}.
	\label{fig:XYZvtang}
	}
\end{figure*}

\subsection{Previous Related Work}

Our focus is on a region of the Galaxy 200 to 500\,pc
from the Sun, above the galactic plane, and spanning galactic
longitudes of $40^\circ$ to $100^\circ$.  Two rich clusters in
this region are the $\delta$\ Lyr cluster
\citep{stephenson_possible_1959} and RSG-5 \citep{roser_nine_2016}.
Each of these clusters was known before Gaia.  Their reported ages are
between 30 and 60\,Myr.  Early empirical evidence that these two
clusters could be part of a large and more diffuse population was
apparent in the Gaia-based photometric analysis of pre-main-sequence
stars by \citet[][compare their Figures~11 and~13 to our
Figure~\ref{fig:XYZvtang}]{Zari2018}.  Further kinematic connections
and complexity were highlighted by \citet{KounkelCovey2019}, who
included these previously known groups in the larger structures dubbed
``Theia~73'' and ``Theia~96''\footnote{See their visualization online
at \url{http://mkounkel.com/mw3d/mw2d.html} (accessed 15 March
2022). Important caveats, particularly for extended groups $\gtrsim 100$\,Myr
old, were presented by \citet{2022arXiv220514160Z}.}.
The connection made by \citet{KounkelCovey2019} between the previously
known open clusters and the other groups in the region was made as
part of an unsupervised clustering analysis of the Gaia DR2 positions
and on-sky velocities with a subsequent manual ``stitching'' step.
Their results support the idea that there is an overdensity of 30 to
60\,Myr old stars in this region of the Galaxy.  \citet{Kerr2021}, in
a volume-limited analysis of the Gaia DR2 point-source catalog out to
one third of a kiloparsec, identified three of the nearest
sub-populations of Cep-Her, dubbed ``Cepheus-Cygnus'', ``Lyra'', and
``Cerberus''.  \citet{Kerr2021} reported ages for each of these
subgroups between 30 and 35 Myr.

\subsection{Member Selection}
\label{subsec:members}

The possibility that the $\delta$~Lyr cluster, RSG-5, and the
sub-populations identified by \citet{Kerr2021} share a common origin
has yet to be fully substantiated, but preliminary clustering results
from the 1\,kpc SPYGLASS analysis (R.~Kerr et al.\ in prep) suggest
the presence of contiguous stellar populations connecting each of
these groups in both space and velocity coordinates.  In other words, the stars
appear to be comoving, though with a continuous gradient in velocity
as a function of position. The lower panels of Figure~\ref{fig:XYZvtang} show this in
detail, where $v_b$ is the distance-corrected proper motion in the
direction of increasing galactic latitude, and $v_{l^*}=v_l \cos b$ is
the distance-corrected proper motion in the direction of increasing
galactic longitude after accounting for the local tangent plane
correction.  Some, but not all, of the gradient in the $v_{l^*}$ vs.~$l$
plane can be understood through a projection effect stemming from the
Sun's motion with respect to the local standard of rest (see also
Figure 11 by \citealt{Zari2018}).  
In this work, our primary interest in this
region of sky stems
from the fact that a portion of it was observed by Kepler
(Figure~\ref{fig:XYZvtang}, top panel).  To further explore this
sub-population, we select candidate Cep-Her members through four
steps, the first three being identical to those described in Section~3
of \citet{Kerr2021}.  We briefly summarize them here.

The first step is to select stars that are photometrically distinct
from the field star population based on Gaia EDR3 magnitudes $\{G,
G_{\rm RP}, G_{\rm BP}\}$, parallaxes and auxiliary reddening
estimates \citep{lallement_gaia-2mass_2019}.  This step yielded \npms\
stars with high-quality photometry and astrometry.  These stars are
either pre-main-sequence K and M dwarfs due to their long contraction
timescales, or massive stars near the zero-age main sequence due to
their rapid evolutionary timescales.

The second step is to perform an unsupervised HDBSCAN clustering on
the photometrically selected population
\citep{campello_hierarchical_2015,mcinnes_hdbscan_2017}.  The
parameters we use in the clustering are $\{ X, Y, Z, c v_b, c v_{l^*}
\} $, where $c$ is
the size-velocity corrective factor, which is taken as $c=6\,{\rm pc /
km\,s}^{-1}$ to ensure that the spatial and velocity scales have
identical standard deviations.  Positions are computed assuming the
\texttt{astropy v4.0} coordinate standard
\citep{astropy_2018}. 
As input
parameters to HDBSCAN, we set the minimum $\epsilon$ threshold past
which clusters cannot be fragmented as $25$\,pc in physical space, and
$25/c$\,\kms\ in velocity.  The minimum cluster size $N$ is set to 10, as
is $k$, the parameter used to define the ``core distance'' density
metric. 
Core distance is the distance to the $k^{\rm th}$ nearest star,
and therefore $k$ acts as a smoothing parameter, where a larger value
reduces the influence of local overdensities smaller than the
scale that interests us.

The unsupervised clustering in this case yielded 8 distinct subgroups.
These groups are then used as the ``seed'' populations, in which the stellar members each have their own
individually-assigned distances to their tenth-nearest
photometrically-young neighbor.  Using those distances, we search the
entire Gaia EDR3 point source catalog for stars that fall within each
star's 10$^{\rm th}$ nearest-neighbor distance.  This third step yields
stars that are spatially and kinematically close to the
photometrically young stars, but which cannot be identified as young
based on their positions in the color--absolute magnitude diagram.

The outcome of the analysis up to the point of the third step is shown
in Figure~\ref{fig:XYZvtang}.  To enable a selection cut that filters
out field-star contaminants, we also compute a weight metric,
$D$, defined such that the group member with the smallest core
distance has $D=1$, the group member with the greatest core
distance has $D=0$, and the weight $D$
scales linearly between the two extremes.  After applying a
set of quality cuts on the astrometry and
photometry\footnote{$\varpi/\sigma_\varpi>5$; $G/\sigma_{G}>50$;
$G_{\rm RP}/\sigma_{G_{RP}}>20$; $G_{\rm BP}/\sigma_{G_{BP}}>20$}
this procedure yields a distribution of weights $D$
that is well described by a log-normal distribution with
$\log_{10}\mathcal{N}(-1.55,0.61)$.
To visualize the results, in Figure~\ref{fig:XYZvtang} we show
12{,}436 objects with $D>0.02$ as gray points, and 
4{,}763 objects with $D>0.10$ as black points.  These thresholds were
selected visually based on the apparent purity with which they yielded
pre-main-sequence stars on a color--absolute magnitude diagram. The
$\delta$~Lyr\ cluster is visible at $(l,b)=(68^\circ,15^\circ$) and $(v_{l^*},
v_b)=(-4.5 ,-4 )\mkms$.  RSG-5 is visible at $(l,b)=(83^\circ,6^\circ)$,
$(v_{l^*}, v_b)=(5.5 ,-3.5 )\mkms$.  Most of the other subclusters, including
in Cep-Cyg ($l,b=90^\circ,7^\circ$) and Cerberus ($l,b=48^\circ,18^\circ$) are
too small or dispersed to have previously been analyzed in great detail.

%
%
Our fourth and final step was to cross-match the candidate Cep-Her
member list against all known Kepler Objects of Interest.  We used the
Cumulative KOI table from the NASA Exoplanet Archive from 27 March
2022, and also compared against the \texttt{q1\_q17\_dr25} table
\citep{thompson_planetary_2018}.  From the candidate members with
weights exceeding 0.02, this yielded 11 known false positives, 6
confirmed planets, and 8 candidate planets (see Appendix~\ref{app:members}).
To determine whether these objects were
potentially consistent with being {\it i)} planets, and {\it ii)}
$\lesssim 10^8$ years old, we inspected the Kepler data validation
reports and Robovetter classifications.  Youth was assessed based on
the presence of rotational modulation at the expected period and
amplitude for stars at least as young as the Pleiades \citep[{e.g.},][]{rebull_rotation_2020}. 
Planetary status was assessed through the Robovetter flags, and by
requiring non-grazing transits with ${\rm S/N}>10$.
Four objects passed both cuts: Kepler-1627, Kepler-1643, KOI-7368, and
KOI-7913.

Figure~\ref{fig:XYZvtang} shows the positions of these KOIs along
various projections.  Kepler-1643 is near the core RSG-5 population
both spatially and kinematically.  KOI-7368 and KOI-7913 are in a
diffuse region $\approx$40\,pc above RSG-5 in $Z$ and $\approx$100\,pc
closer to the Sun in $Y$.  In tangential galactic velocity space,
there is some kinematic overlap between the region containing the latter two KOIs
and the main RSG-5 group.

%
%
We define two sets of stars in the local vicinity of our objects of
interest.  For candidate RSG-5 members near Kepler-1643, we require:
\begin{align}
  X/{\rm pc} &\in [45, 75] \nonumber \\
  Y/{\rm pc} &\in [320, 350] \nonumber \\
  Z/{\rm pc} &\in [40, 80] \nonumber \\
  v_b/{\rm km\,s^{-1}} &\in [-4, -2.5] \nonumber \\
  v_{l^*}/{\rm km\,s^{-1}} &\in [3.5, 6], \nonumber
\end{align}
though RSG-5 does have a greater spatial extent toward smaller
$X$ (Figure~\ref{fig:XYZvtang}, middle panels).  For the diffuse stars
near KOI-7368 and KOI-7913, we require
\begin{align}
  X/{\rm pc} &\in [20, 70] \nonumber \\
  Y/{\rm pc} &\in [230, 270] \nonumber \\
  Z/{\rm pc} &\in [75, 105] \nonumber \\
  v_b/{\rm km\,s^{-1}} &\in [-3.5, -1.5] \nonumber \\
  v_{l^*}/{\rm km\,s^{-1}} &\in [2, 6] \nonumber
\end{align}
and we call this latter set of stars ``CH-2'', using the preliminary 
Cep-Her (CH) subgroup identifier from R.~Kerr et al.\ (in prep).
These cuts yielded
\nrsgfive\ candidate RSG-5 members, and \nchtwo\ candidate CH-2
members.  These stars are listed in Appendix~\ref{app:members}, as is the set of Cep-Her
candidates that was observed by Kepler.

\subsection{The Cluster's Age}
\label{sec:clusterage}

\begin{figure*}[tp]
	\begin{center}
		\leavevmode
		\subfloat{
			\includegraphics[width=0.49\textwidth]{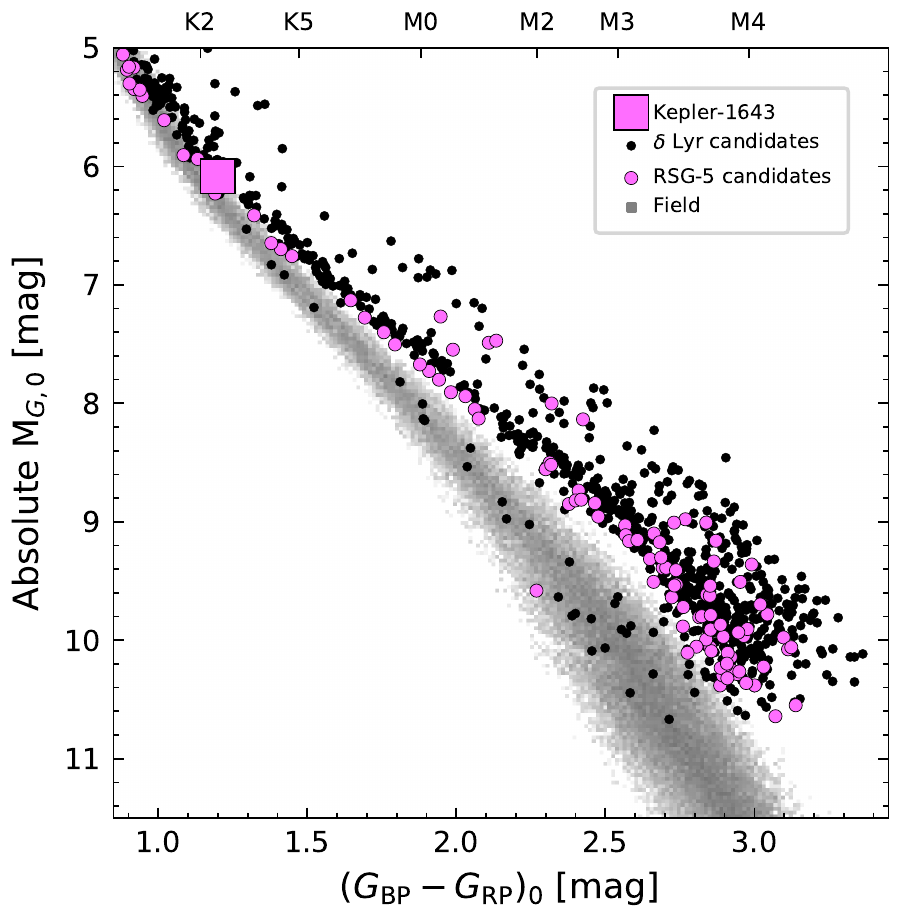}
			\includegraphics[width=0.49\textwidth]{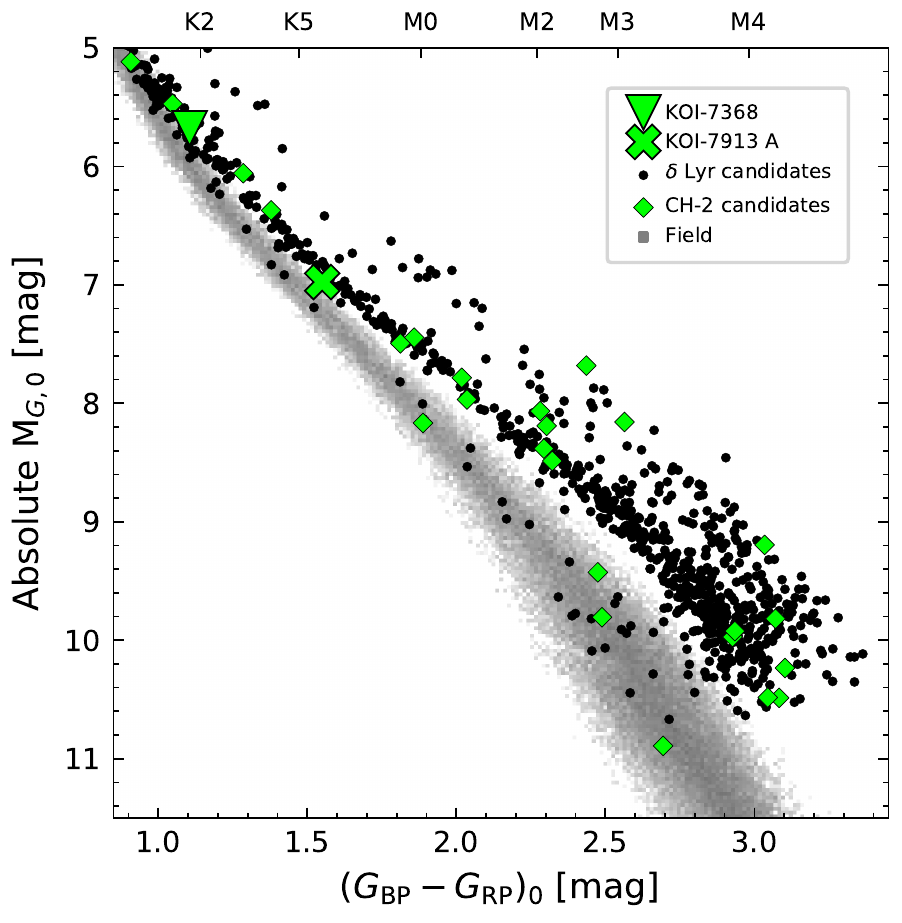}
		}
		
		\vspace{-0cm}
		\subfloat{
			\includegraphics[width=0.49\textwidth]{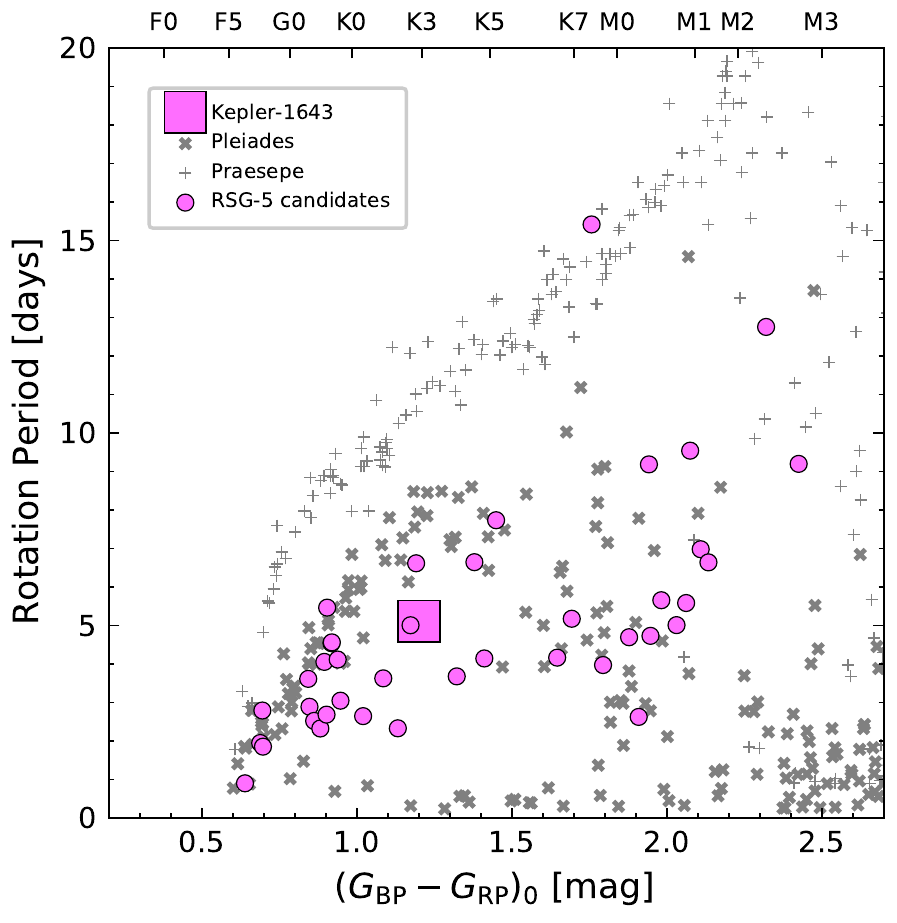}
			\includegraphics[width=0.49\textwidth]{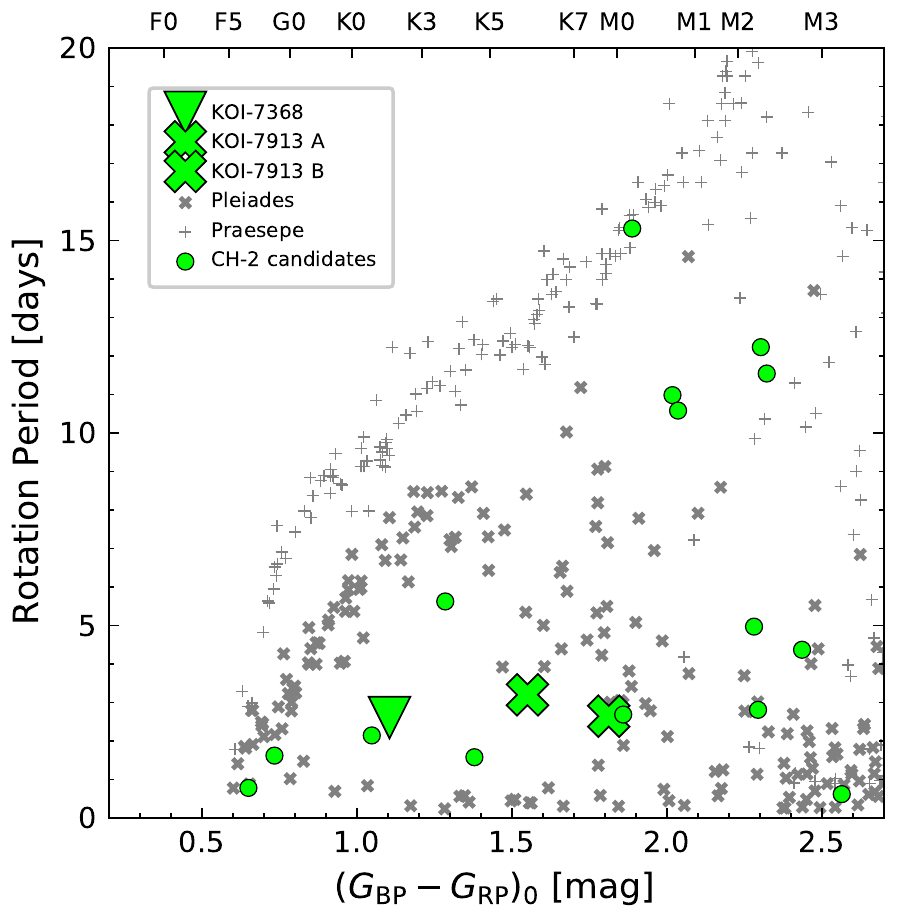}
		}
	\end{center}
	\vspace{-0.5cm}
	\caption{
    {\bf Age-diagnostic diagrams from the stellar groups near
    Kepler-1643, KOI-7368, and KOI-7913.} 
    {\it Top row}: Color--absolute magnitude diagram of candidate
    Cep-Her members, plotted over candidate members of the
    $\delta$~Lyr~cluster ($\approx38$\,Myr;
    \citealt{bouma_kep1627_2022}) and the Gaia EDR3 Catalog of Nearby
    Stars (gray background).  The left and right columns shows stars
    in RSG-5 and CH-2, respectively.  The range of colors is truncated
    to emphasize the pre-main-sequence; approximate spectral types are shown on the upper axes.  Stars that fall far below the
    cluster sequences are field interlopers.  {\it Bottom row}:
    TESS and ZTF-derived stellar rotation periods, with the Pleiades
    ($\approx 112$\,Myr) and Praesepe ($\approx 650$\,Myr) shown for
    reference \citep{rebull_rotation_2016a,douglas_poking_2017}.  The
    detection efficiency for reliable rotation periods falls off
    beyond $\bpmrpo \gtrsim 2.6$.
	\label{fig:age}
	}
\end{figure*}

\subsubsection{Color--Absolute Magnitude Diagram}
\label{sec:camd}

Color--absolute magnitude diagrams (CAMDs) of the candidate RSG-5 and
CH-2 members are shown in the upper row of Figure~\ref{fig:age}.  The
stars from the $\delta$~Lyr cluster are from
\citet{bouma_kep1627_2022}, and the field stars are from the Gaia EDR3
Catalog of Nearby Stars \citep{gaia_gcns_2021}.  To make these
diagrams, we imposed the data filtering criteria from
\citet[][Appendix~B]{GaiaCollaboration2018}, which include binaries
while omitting artifacts from for instance low
photometric signal to noise, or a small number of visibility periods.
We then corrected for extinction using the
\citet{lallement_threedimensional_2018} dust
maps and the extinction
coefficients $k_X\equiv A_X/A_0$ from \citet{GaiaCollaboration2018},
assuming that $A_0 = 3.1 E(B-V)$.  This yielded a mean and standard
deviation for the reddening of $E(B-V)=0.036\pm0.002$ for RSG-5, and
$E(B-V)=0.017\pm0.001$ for CH-2.  By way of comparison, in
\citet{bouma_kep1627_2022} the same query for the $\delta$~Lyr cluster
yielded $E(B-V)=0.032\pm0.006$.  Finally, for the plots we set the
color axis to best visualize the region of maximal age information
content: the pre-main-sequence.

The CAMDs show that for RSG-5, all but one of the candidate
members are on a tight pre-main-sequence locus.  Quantitatively,
88/89 stars with $\bpmrpo \geq 1.5$ are consistent with being on the
pre-main-sequence. This implies a false
positive rate of a few percent, at most.  In comparison, our
reference
sample (the $\delta$~Lyr candidates) has a false positive rate of
$\approx$12\%, based on the number of stars that photometrically
overlap with the field population.  For CH-2, our membership selection gives 27
objects in the color range displayed, and 23 of them appear to be
consistent with being on the pre-main-sequence.  This
would imply a false
positive rate in CH-2 of $\approx$15\%.

Figure~\ref{fig:age} also shows that most RSG-5 and CH-2 members
overlap with the $\delta$~Lyr cluster on the CAMD, and that the groups are
therefore roughly the same age.  To quantify this, we use the method
introduced by \citet[][their Section~6.3]{gagne_mutau_2020}.  The idea
is to fit the pre-main-sequence loci of a set of reference clusters,
and to then model the locus of the target cluster as a linear
combination of these reference cluster loci.  For our reference
clusters, we used UCL, IC\,2602, and the Pleiades, with the
memberships reported by \citet{Damiani2019} and
\cite{CantatGaudin2018a} respectively.  We adopted ages of 16\,Myr for
UCL \citep{pecaut_star_2016}, 38\,Myr for IC\,2602
\citep{david_ages_2015,randich_gaiaeso_2018} and 112\,Myr for the
Pleiades \citep{dahm_2015}.  These assumptions and the subsequent
processing steps taken to exclude field stars and binaries were
identical to those described in \citet{bouma_kep1627_2022}.  The mean
and uncertainty of the resulting age posterior are $46^{+9}_{-7}$\,Myr
for RSG-5, and $36^{+10}_{-8}$\,Myr for CH-2.  For comparison, this
procedure yields an age for the $\delta$~Lyr cluster of
$38^{+6}_{-5}$\,Myr.  The older isochronal age of RSG-5 is consistent
with its location relative to the $\delta$~Lyr cluster in the upper
left panel of Figure~\ref{fig:age}.
Generally speaking, this method is expected to be accurate provided
that the metallicities of IC\,2602 and the Cep-Her groups (RSG-5,
CH-2, and the $\delta$~Lyr cluster) are roughly identical.  The
spectroscopic metallicities that we find in Section~\ref{sec:stars}
suggest that this is indeed the case.
While in reality stellar populations do not evolve linearly in
the dimensions of absolute magnitude versus color, in our case the
Cep-Her loci are nearly indistinguishable from IC\,2602 (e.g.,
Figure~3 of \citealt{bouma_kep1627_2022}).  Systematic errors incurred
in the age from the non-linear evolution are therefore likely much
smaller than the $\approx$10\,Myr systematic uncertainty in the
absolute reference age for IC\,2602 itself
\citep{david_ages_2015,randich_gaiaeso_2018}.

\subsubsection{Stellar Rotation Periods}
\label{sec:rotation}

An independent way to assess the age of the candidate cluster members
is to measure their stellar rotation periods.  This approach can be
achieved using surveys such as TESS \citep{ricker_transiting_2015} and
the Zwicky Transient Facility (ZTF, \citealt{bellm_zwicky_2019}); it
leverages a storied tradition of measuring rotation periods of stars
in benchmark open clusters \citep[see
{e.g.},][]{skumanich_time_1972,curtis_rup147_2020}.  The TESS data in
our case are especially useful, since they provide 3 to 5 lunar months
of photometry for all of our candidate CH-2 and RSG-5 members.

We selected stars suitable for gyrochronology by requiring 
$\bpmrpo \geq 0.6$ to focus on FGKM stars that experience magnetic
braking.
For TESS, we also restricted our sample to $G<16$, to ensure the stars
are bright enough to extract usable light curves from the full-frame
images.  The magnitude cut corresponds to $\bpmrpo < 2.6$
($\sim$M3V) at the relevant distances.  These cuts gave 19 stars in
CH-2 and 42 stars in RSG-5.  We extracted light curves from the TESS
images using the \texttt{unpopular} package \citep{hattorio_2021_cpm},
and regressed them against systematics with its causal pixel model.
We measured rotation periods using Lomb-Scargle periodograms and
visually vetted the results using an interactive program that allows
us to switch between TESS Cycles, select particular sectors, flag
stars with multiple periods, and correct half-period harmonics. For
ZTF, we used the same color cut to focus on FGKM stars, but restricted
the sample to $13 < G < 18$ to avoid the saturation limit on the
bright end and ensure sufficient photometric precision at the faint
end. We followed the procedure outlined in \citet{curtis_rup147_2020}:
we downloaded $8'\times8'$ image cutouts, ran aperture photometry for
the target and neighboring stars identified with Gaia, and used them
to define a systematics correction to refine the target light curves. 

The lower panels of Figure~\ref{fig:age} show the results.  In RSG-5,
39/42 stars have rotation periods at least as fast as the Pleiades (93\%).  This
numerator omits the two stars with periods $>$$12$\,days visible in
the lower-left panel of Figure~\ref{fig:age}.  The age interpretation
for these latter stars, particularly the $\approx$M2.5 dwarf, is not
obvious.  \citet{rebull_usco_2018} for instance have found numerous
M-dwarfs with 10-12 day rotation periods at ages of USco
($\sim$$8$\,Myr), and some  still exist at the age of
LCC ($\sim$$16$\,Myr; \citealt{2022AJ....164...80R}).  Regardless, since
only one field star
outlier seems to be present on the RSG-5 CAMD,
the fact
that we do not detect rotation periods for $\approx$7\%
of stars should perhaps be taken as an indication for the fraction of
stars for which rotation periods might not be detectable, due to
{e.g.}, pole-on stars having lower amplitude starspot modulation.
Field star contamination is another possible contributor.

For CH-2, 13/19 stars have rotation periods that are obviously faster
than their counterparts in the Pleiades.  4 stars, not included
in the preceding numerator, are M-dwarfs with rotation periods between
10 and 12.5 days.  As previously noted, the age interpretation for
these M-dwarfs is ambiguous.  If none are cluster members,
the rotation period detection fraction is 68\%; if all are members, it
is 89\%.

This sets an upper bound on the contamination fraction in our candidate
CH-2 members at about one in three.  Combined with the roughly one in
six contaminant rate implied by the earlier CAMD analysis, this
suggests that the sample of candidate CH-2 members is more polluted by
field stars than the RSG-5 sample.

It is challenging to convert these stellar rotation periods to a
precise age estimate, since on the pre-main-sequence the stars are
spinning up due to thermal contraction rather than down due to
magnetized braking.  Regardless, the rotation period distributions of
both CH-2 and RSG-5 seem consistent with other 30\,Myr to 50\,Myr
clusters ({e.g.}, IC\,2602 and IC\,2391;
\citealt{douglas_stephanie_t_2021_5131306}).  They also seem
consistent with the false positive rates estimated from the
color--absolute magnitude diagrams.

\section{The Stars}
\label{sec:stars}

\begin{deluxetable}{lccc}
\tabletypesize{\scriptsize}
\tablecaption{Selected system parameters of Kepler-1643, KOI-7368, and KOI-7913. \label{tab:sysparams}}
\tablenum{1}

\tablehead{
\colhead{Parameter} & \colhead{Value} & \colhead{Uncertainty} & \colhead{Comment}
}

\startdata
\hline
\multicolumn{4}{c}{\emph{Kepler-1643}} \\
\hline
{\it Stellar parameters:} & & & \\
  Gaia $G$~[mag]                             & $13.836$           & $\pm 0.003$                & A \\
  $T_{\rm eff}$~[K]                          & $4916$             & $\pm 110$                  & B \\
  $\log g_\star$~[cgs]                       & $4.502$            & $\pm 0.035$                & C \\
  $R_\star$~[R$_{\odot}$]                    & $0.855$            & $\pm 0.044$                & C \\
  $M_\star$~[M$_{\odot}$]                    & $0.845$            & $\pm 0.025$                & C \\
  $\rho_\star$~[g~cm$^{-3}$]                 & $1.910$            & $\pm 0.271$                & C \\
  $P_{\rm rot}$~[days]                       & $5.106$            & $\pm 0.044$                & D \\
  Li EW~[m\AA]                               & $130$              & $+6$, $-5$                 & E \\
{\it Transit parameters:} & & & \\
  $P$~[days]                                 & $5.3426258$        & $\pm 0.0000101$            & D \\
  $R_{\rm p}/R_\star$                        & $0.025$            & $\pm 0.001$                & D \\
  $b$                                        & $0.58$             & $\pm 0.05$                 & D \\
  $R_{\rm p}$~[R$_{\oplus}$]                 & $2.32 $            & $\pm 0.14$                 & D \\
  $t_{14}$~[hours]                           & $2.41$            & $\pm 0.06$                & D \\
\hline
\multicolumn{4}{c}{\emph{KOI-7368}} \\
\hline
{\it Stellar parameters:} & & & \\
  Gaia $G$~[mag]                             & $12.831$           & $\pm 0.004$                & A \\
  $T_{\rm eff}$~[K]                          & $5241$             & $\pm 100$                   & F \\
  $\log g_\star$~[cgs]                       & $4.499$            & $\pm 0.030$                & C \\
  $R_\star$~[R$_{\odot}$]                    & $0.876$            & $\pm 0.035$                & C \\
  $M_\star$~[M$_{\odot}$]                    & $0.879$            & $\pm 0.018$                & C \\
  $\rho_\star$~[g~cm$^{-3}$]                 & $1.840$            & $\pm 0.225$                & C \\
  $P_{\rm rot}$~[days]                       & $2.606$            & $\pm 0.038$                & D \\
  Li EW~[m\AA]                               & $236$              & $+16$, $-14$               & E \\
{\it Transit parameters:} & & & \\
  $P$~[days]                                 & $6.8430341$        & $\pm 0.0000125$            & D \\
  $R_{\rm p}/R_\star$                        & $0.023$            & $\pm 0.01$                 & D \\
  $b$                                        & $0.50$             & $\pm 0.06$                 & D \\
  $R_{\rm p}$~[R$_{\oplus}$]                 & $2.22$             & $\pm 0.12$                 & D \\
  $t_{14}$~[hours]                           & $2.79 $           & $\pm 0.07$                & D \\
\hline
\multicolumn{4}{c}{\emph{KOI-7913}} \\
\hline
{\it Stellar parameters:} & & & \\
  Gaia $G$~[mag]                             & $14.200$           & $\pm 0.003$                & A \\
  $T_{\rm eff,A}$~[K]                        & $4324$             & $\pm 70$                   & B \\
  $T_{\rm eff,B}$~[K]                        & $4038$             & $\pm 70$                   & B \\
  $\log g_{\star,A}$~[cgs]                   & $4.523$            & $\pm 0.043$                & C \\
  $R_{\star,A}$~[R$_{\odot}$]                & $0.790$            & $\pm 0.049$                & C \\
  $M_{\star,A}$~[M$_{\odot}$]                & $0.760$            & $\pm 0.025$                & C \\
  $\rho_{\star,A}$~[g~cm$^{-3}$]             & $2.172$            & $\pm 0.379$                & C \\
  $P_{\rm rot,A}$~[days]                     & $3.387$            & $\pm 0.016$                & D \\
  $P_{\rm rot,B}$~[days]                     & $2.642$            & $\pm 0.067$                & D \\
  (Li EW)$_{\rm A}$~[m\AA]                   & $65$               & $+8$, $-6$                 & E \\
  (Li EW)$_{\rm B}$~[m\AA]                   & $42$               & $+12$, $-19$               & E \\
  $\Delta G_{\rm AB}$~[mag]                  & $0.51$             & $\pm 0.01$                 & G \\
  Apparent sep.~[au]                   		   & $959.4$            & $\pm 1.9$                  & G \\
{\it Transit parameters$^\dagger$:} & & & \\
  $P$~[days]                                 & $24.278571$       & $\pm 0.000263$            & D \\
  $R_{\rm p}/R_\star$                        & $0.027$            & $\pm 0.001$                & D \\
  $b$                                        & $0.30$             & $\pm 0.15$                 & D \\
  $R_{\rm p}$~[R$_{\oplus}$]                 & $2.34$             & $\pm 0.18$                 & D \\
  $t_{14}$~[hours]                           & $4.40$            & $0.21$                    & D \\
\enddata
\tablecomments{
  $^\dagger$The planet orbits KOI-7913 A (Section~\ref{subsec:validation}).
  (A) \citet{gaia_collaboration_2021_edr3}.
  (B) HIRES SpecMatch-Emp \citep{yee_SM_2017}.
  (C) Cluster isochrone \citep{choi_mesa_2016, bressan_parsec_2012}.
  (D) Kepler light curve.
  (E) HIRES/TRES \citep{bouma_2021_ngc2516}.
  (F) TRES SPC \citep{Buchhave2012,2021tsc2.confE.124B}.
  (G) Magnitude difference and apparent physical separation between primary and secondary; from Gaia EDR3.
  (H) HIRES SpecMatch-Synth \citep{petigura_cksi_2017}.
}
\end{deluxetable}

Many of the salient properties of the Kepler objects of interest in Cep-Her can be
gleaned from Figure~\ref{fig:age}.  The stars span spectral types of
G8V (Kepler-1627) to K6V (KOI-7913 A).  The secondary in the KOI-7913
system has spectral type $\approx$K8V.  And since a star with
Solar mass and metallicity arrives at the zero-age main sequence at
$t\approx40$ Myr \citep{choi_mesa_2016}, these stars are all in the
late stages of their pre-main-sequence contraction.  

The adopted stellar parameters are listed in
Table~\ref{tab:sysparams}.  The stellar surface gravity, radius, mass,
and density are found by interpolating against the MIST isochrones
in reddening-corrected absolute $G$-band magnitude as a function of $\bpmrpo$\ color
\citep{choi_mesa_2016}.  The statistical uncertainties from this
technique mostly originate from the parallax uncertainties; the
systematic uncertainties are taken to be the absolute difference
between the PARSEC \citep{bressan_parsec_2012} and MIST isochrones.
Reported uncertainties are a quadrature sum of the statistical and
systematic components. 

To verify these parameters, determine the stellar effective temperatures, and to analyze youth
proxies such as the Li 6708\,\AA\ doublet and H$\alpha$, we acquired
high resolution optical  spectra.  We also acquired high resolution imaging for each system, to
constrain the existence of visual companions, including possible bound
binaries.  We give the system-by-system details in Sections~\ref{subsec:kep1643}
through~\ref{subsec:koi7913}, and summarize their implications for
the youth of the stars in Section~\ref{subsec:specyouth}.

\subsection{Kepler\,1643}
\label{subsec:kep1643}

\paragraph{Spectra}
For Kepler-1643, we acquired two iodine-free spectra from Keck/HIRES
on the nights of 2020 Aug 16 and 2021 Oct 25.  The acquisition and
analysis followed the usual techniques of the California Planet Survey
\citep{howard_cps_2010}.  We derived the stellar parameters ($T_{\rm
eff}, \log g, R_\star$) using \texttt{SpecMatch-Emp}
\citep{yee_SM_2017}, which yielded values in $<$$1$-$\sigma$ agreement
with those from the cluster-isochrone method.  This approach also
yielded $[{\rm Fe/H}]=0.13 \pm 0.09$.  Using the broadened synthetic
templates\footnote{The broadening is calculated using the joint
rotational and macroturbulent broadening kernel from \citet{hirano_2011},
assuming that the macroturbulent velocity scales with effective temperature similar
to the prescription from \citet{2014MNRAS.444.3592D}.  The latter assumption could be a 
source of systematic uncertainty in our equatorial velocity measurements, since
the macroturbulent velocity could be systematically higher (or lower) on
the pre-main-sequence than it is for more slowly rotating field stars.}
 from \texttt{SpecMatch-Synth} \citep{petigura_cksi_2017}, we
found $v\sin i = 9.3 \pm 1.0\,\mkms$.  The systemic radial velocity at
the two epochs was $-9.1 \pm 1.9\,\mkms$ and $-7.8\pm 1.2\,\mkms$
respectively, and was calculated
following the methods of \citet{chubak_2012}.  To infer the equivalent width of
the \ion{Li}{1} 6708\,\AA\ doublet, we followed the procedure
described by \citet{bouma_2021_ngc2516}. In brief, this
involved calculating the line width by numerically integrating a
single best-fit Gaussian over a local window, and estimating the
uncertainties through a Monte Carlo procedure in which the continuum
normalization was allowed to vary through a bootstrap approach based
on the local scatter in the spectra.  For Kepler-1643, this yielded a
strong detection: ${\rm EW}_{\rm Li} = 130^{+6}_{-5}$\,m\AA, with
values consistent at $<$$1$-$\sigma$ between the two epochs.   The
quoted value does not correct for the \ion{Fe}{1} blend at
6707.44\,\AA.  Given the purported age and effective temperature of
the star, the lithium equivalent width is somewhat low.  We discuss
this in greater depth in Section~\ref{subsec:specyouth}.

\paragraph{High-Resolution Imaging}
We acquired adaptive optics imaging of Kepler-1643 on the night of
2019 June 28 using the NIRC2 imager on Keck-II.  Using the narrow
camera (FOV = 10.2\arcsec), we obtained 4 images in the $K'$ filter
($\lambda = 2.12\,\mu$m) with a total exposure time of 320\,s. 
The images did not show any additional visual companions.
We analyzed the data following \citet{kraus_impact_2016}, and
determined the detection limits by analyzing the residuals after
subtracting an empirical PSF template.  This procedure yielded
contrast limits of $\Delta K' = 4.1$ mag at $\rho = 150$ mas, $\Delta
K' = 5.8$ mag at $\rho = 300$ mas, and $\Delta K' = 8.3$ mag
at $\rho > 1000$ mas.

\subsection{KOI-7368}
\paragraph{Spectra}
For KOI-7368, we acquired a spectrum on 2015 June 1 using the echelle
spectrograph (TRES; \citealt{furesz_tres_2008}) mounted at the
Tillinghast 1.5m at the Fred Lawrence Whipple Observatory.  The
Stellar Parameter Classification pipeline for TRES has been described
by \citet{2021tsc2.confE.124B}.  It is based on the synthetic template
library constructed by \citet{Buchhave2012}.  The resulting stellar
parameters ($T_{\rm eff}, \log g, R_\star$) agreed with those from the
cluster-isochrone method within $1$-$\sigma$.  Auxiliary
spectroscopic parameters included the metallicity $[{\rm Fe/H}]= -0.02
\pm 0.08$, the equatorial velocity $v\sin i = 20.2 \pm 1.0\,\mkms$,
and the systemic velocity ${\rm RV}_{\rm sys} = -10.9 \pm 0.2\,\mkms$.
The Li 6708\AA\ EW measurement procedure yielded ${\rm EW}_{\rm Li} =
236^{+16}_{-14}$\,m\AA.

\paragraph{High-Resolution Imaging}
We acquired adaptive optics imaging of KOI-7368 on the night of 2019
June 12, again using NIRC2.  The observational configuration and
reduction were identical as for Kepler-1643.  No companions were
detected, and the analysis of the image residuals yielded contrast
limits of $\Delta K' = 5.2$ mag at $\rho = 150$ mas, $\Delta K' = 6.7$
mag at $\rho = 300$ mas, and $\Delta K' = 8.7$ mag at $\rho > 1000$
mas.

\subsection{KOI-7913}
\label{subsec:koi7913}

\paragraph{Binarity}
KOI-7913 is a binary.  The north-west primary is
$\approx$0.5 magnitudes brighter than the south-east secondary in
optical passbands.  The two stars are separated in Gaia EDR3 by
$3\farcs5$ on-sky, and have parallaxes consistent within $1$-$\sigma$
(with an average $\varpi=3.66 \pm 0.01$\,mas).  The apparent on-sky
separation is $959 \pm 2$ au.  The Gaia EDR3 proper motions are also
very similar.  Since the two stars were resolved in the Kepler Input
Catalog and are roughly one Kepler pixel apart, an accurate crowding
metric has already been applied in the NASA Ames data products to
correct the mean flux level \citep{2017ksci.rept....6M}.  This is
important for deriving accurate transit depths.

\paragraph{Spectra}
We acquired Keck/HIRES spectra for KOI-7913 A on the night of 2021 Nov
13, and KOI-7913 B on the night of 2021 Oct 26.  The
\texttt{SpecMatch-Emp} machinery yielded $T_{\rm eff,A} = 4324 \pm
70\,{\rm K}$, and $T_{\rm eff,B} = 4038 \pm 70\,{\rm K}$.  These temperatures as well as the other spectroscopic parameters agreed with those from the cluster
isochrone method within 1-$\sigma$.  For the primary, we also found $[{\rm Fe/H}]= -0.06 \pm
0.09$, $v\sin i = 13.3 \pm 1.0\,\mkms$, and ${\rm RV}_{\rm sys} =
-17.8 \pm 1.1\,\mkms$.  For the secondary, these same parameters were
$[{\rm Fe/H}]= -0.01 \pm 0.09$, $v\sin i = 10.7 \pm 1.0\,\mkms$, and
${\rm RV}_{\rm sys} = -18.8 \pm 1.1\,\mkms$.  The primary showed lithium in absorption
with ${\rm EW}_{\rm Li} =
65^{+8}_{-6}$\,m\AA, while the secondary had a marginal detection of ${\rm EW}_{\rm Li} =
42^{+12}_{-19}$\,m\AA. .  Both components displayed 
H$\alpha$ in emission.  Given the spectral types of the stars, these observations
are consistent with a $\approx$40 Myr age for KOI-7913 (see
Section~\ref{subsec:specyouth}).

\paragraph{High-Resolution Imaging}
We acquired adaptive optics imaging of KOI-7913 on the night of 2020
Aug 27 using the NIRC2 imager.  The observational configuration and
reduction were identical as before.  The images showed KOI-7913 A,
KOI-7913 B, and an additional faint neighbor $\approx$$0\farcs99$ due East
of KOI-7913 B.  Applying the PSF-fitting routines from
\citet{kraus_impact_2016}, the tertiary object has a separation $\rho =
4397 \pm 3$\,mas from the primary, at a position angle $231.17^\circ
\pm 0.02^\circ$, with $\Delta K' = 6.97 \pm 0.04$.  While it is too
faint to affect the interpretation of the transit signal, it would be
amusing if this faint neighbor were comoving and therefore part of the system, 
because it would have a mass between 10 and
15\,M$_{\rm Jup}$ at an assumed age of 40\,Myr.  Additional imaging epochs will tell.

\subsection{Spectroscopic Youth Indicators}
\label{subsec:specyouth}

\begin{figure*}[t]
	\begin{center}
		\leavevmode
			\includegraphics[width=0.9\textwidth]{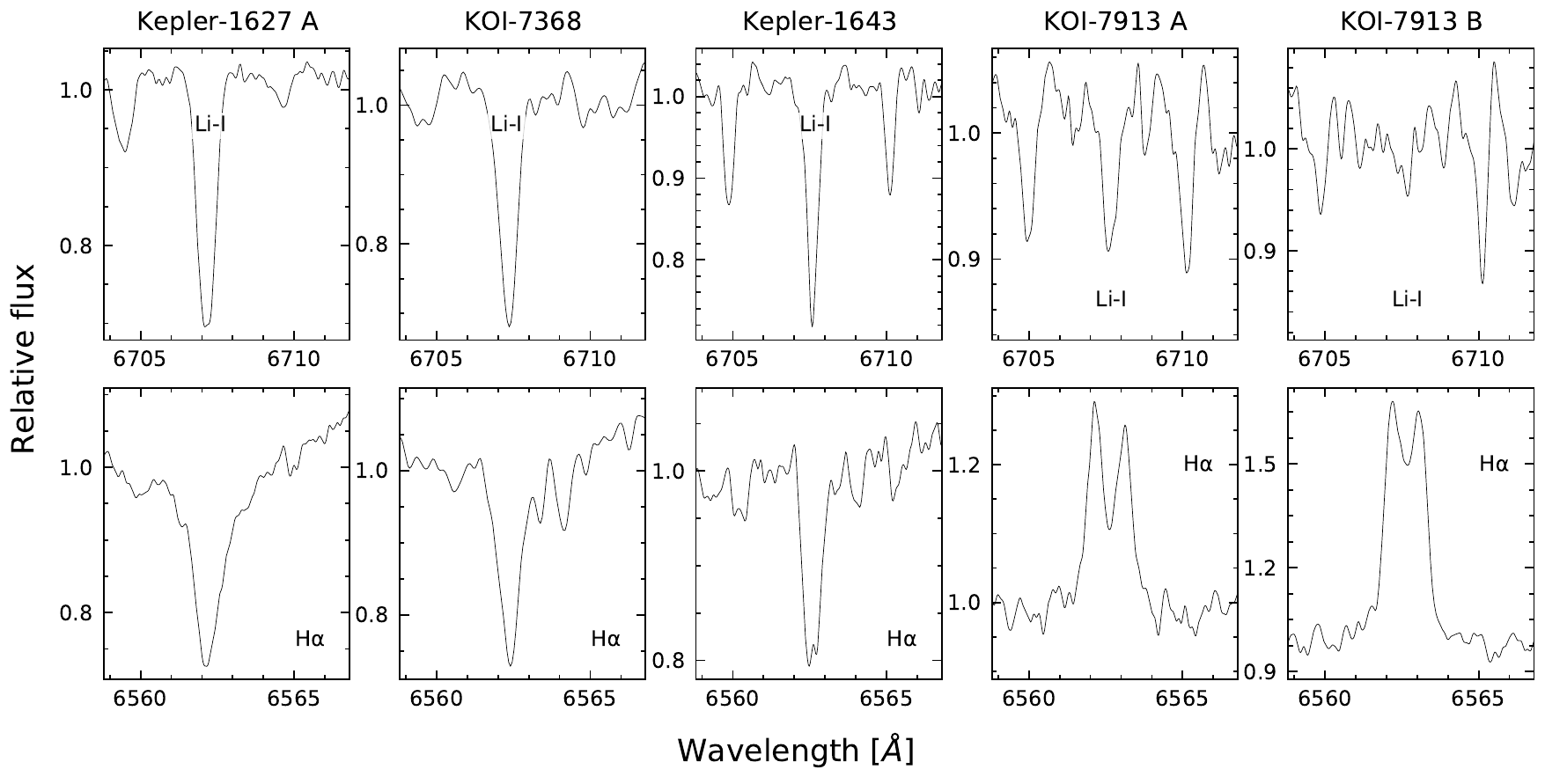}
	\end{center}
	\vspace{-0.5cm}
	\caption{
    {\bf Spectroscopic youth diagnostics for Kepler-1627, KOI-7368,
    Kepler-1643, and KOI-7913 AB. }
    The spectra are shown in the observed frame, and the stars are
    sorted left-to-right in order of decreasing effective temperature.
    \label{fig:koiyouthindicators}
	}
\end{figure*}

Figure~\ref{fig:koiyouthindicators} shows key portions of the HIRES
and TRES spectra for the Kepler objects in Cep-Her.  Lithium
absorption is obvious at 6708\AA\ in all stars except KOI-7913 B.  
H$\alpha$ is in emission for both components of
KOI-7913, and in absorption for the hotter stars.  Here, we compare
these observations against benchmark open clusters in order to
assess their implications for the stellar ages.

\subsubsection{Lithium}

\begin{figure}[tp]
	\begin{center}
		\leavevmode
		\subfloat{
			\includegraphics[width=0.47\textwidth]{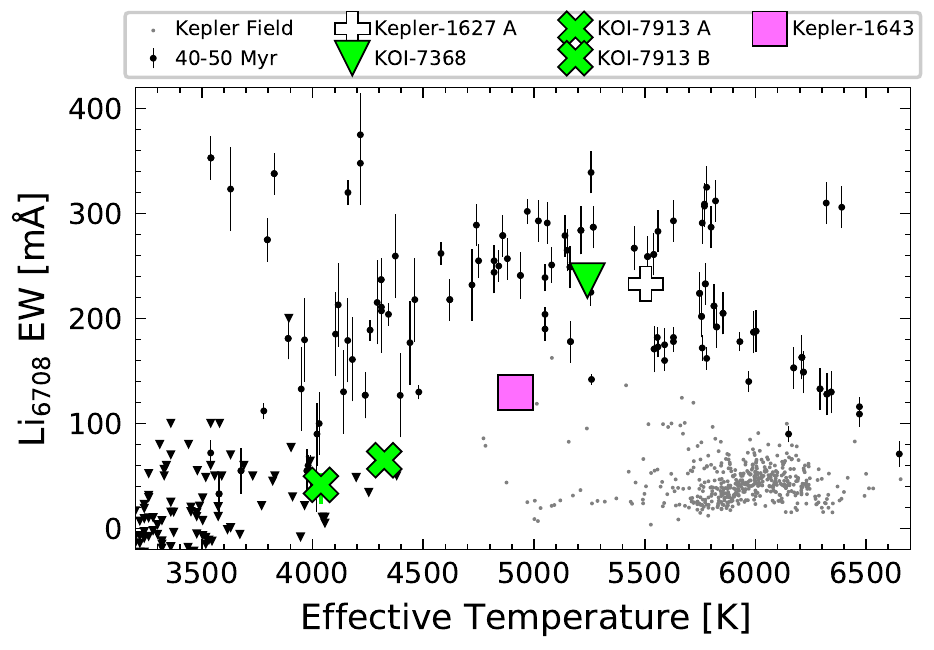}
		}
	
		\vspace{-0.35cm}
		\subfloat{
			\includegraphics[width=0.47\textwidth]{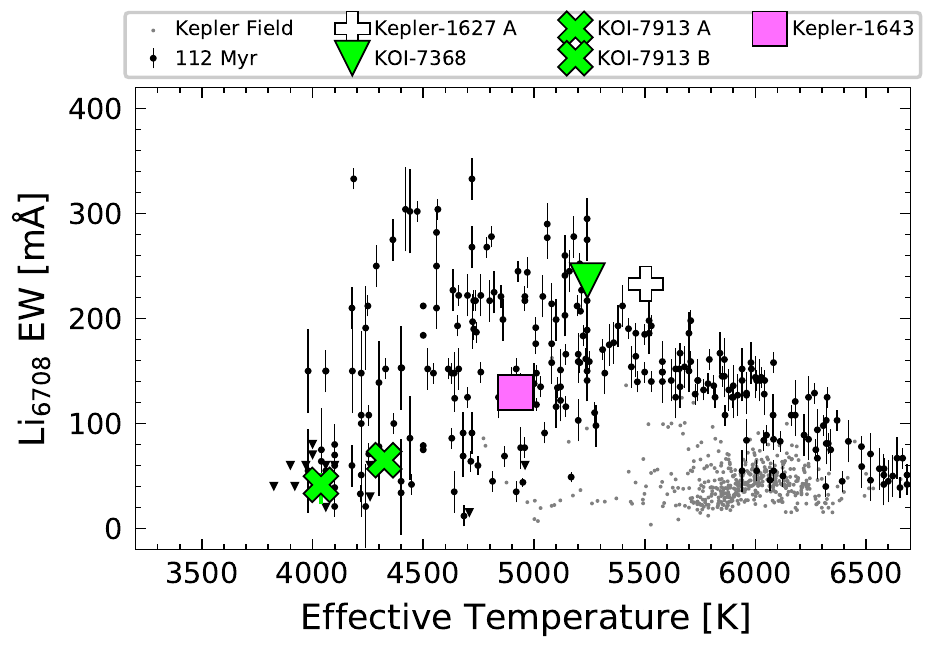}
		}
	
		\vspace{-0.35cm}
		\subfloat{
			\includegraphics[width=0.47\textwidth]{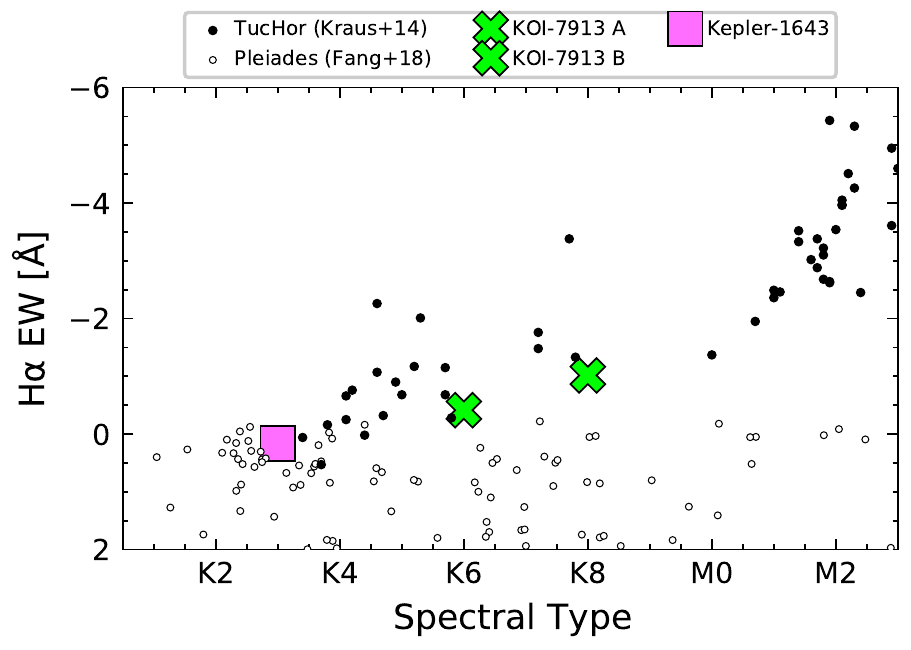}
		}
	\end{center}
	\vspace{-0.6cm}
	\caption{
    {\bf Lithium 6708\AA\ and H$\alpha$ equivalent widths for the
    objects of interest compared to young open clusters and field
    stars. } Positive equivalent width means absorption; negative
    equivalent width means emission.  
    {\it Top \& middle}:
    The field stars are KOIs from
    \citet{berger_identifying_2018}.  The ``40-50 Myr'' reference
    stars ({\it left}) are from IC\,2602 \citep{randich_2001} and
    Tuc-Hor \citep{kraus_stellar_2014}.  The ``112 Myr'' stars
    are from the Pleiades
    \citep{soderblom_evolution_1993,jones_evolution_1996,bouvier_pleiades_lirot_2018}.
    The statistical uncertainties on
    the equivalent widths are shown, or else are
    smaller than the markers.
    {\it Bottom}:
    The H$\alpha$ comparison is against Tuc-Hor
    \citep[$\approx$$40$\,Myr;][]{kraus_stellar_2014} and the Pleiades \citep{fang_2018}.
    \label{fig:lithium}
	}
\end{figure}

Figure~\ref{fig:lithium} compares the measured lithium equivalent
widths of the Kepler objects against a few reference populations.  We
selected reference studies from the literature only when upper limits were
explicitly reported.  KOI-7368 and KOI-7913~A have secure lithium
detections, while for KOI-7913~B the detection is marginal (${\rm EW}_{\rm
Li} = 42^{+12}_{-19}$\,m\AA).  For all three stars, as well as for
Kepler-1627~A, the observed lithium equivalent width is consistent
with the stellar effective temperatures and a $\approx40$\,Myr age.

Kepler-1643, in RSG-5, is conspicuously below the 40-50$\,$Myr
sequence in the top panel of Figure~\ref{fig:lithium}, though
above the field stars (${\rm EW}_{\rm Li} = 130^{+6}_{-5}$\,m\AA).

Quantitatively, there are 14 reference stars within $\pm$150\,K of Kepler-1643.
The mean and standard deviation of their lithium EWs is $255 \pm 31$\,m\AA,
which implies that Kepler-1643 is $4.0$-$\sigma$ discrepant from expectations.

The middle panel shows a comparison against the Pleiades, where
Kepler-1643 is more consistent with the observed dispersion in
lithium.

One explanation for the low Li equivalent width
in Kepler-1643 relative to the comparison stars could be that it is a field
interloper; another could be that RSG-5 is much older than 50\,Myr.  We do
not favor either explanation.  
RSG-5 cannot be much older than 50\,Myr
based on its proximity to the $\delta$~Lyr cluster and IC\,2602 in the
CAMD, and because it is below the Pleiades in the 
rotation versus color diagram (Figure~\ref{fig:age}).
Kepler-1643 also seems highly unlikely to be a field interloper, 
because we demonstrated a few-percent false positive probability
in our spatio-kinematic selection of RSG-5 members, and there is a 
similar independent chance
($\approx$$1\%$) of a field K2V star having a rotation period below the
Pleiades \citep{mcquillan_rotation_2014}.
This yields a puzzle: how could a star have spatial, kinematic, and rotational
evidence consistent with being in a $\approx$50\,Myr cluster,
but a low lithium content?

Our preferred explanation for Kepler-1643's meager lithium 
content is that the reference samples of IC\,2602 and Tuc-Hor
stars may not fully explore all possible lithium equivalent widths at this age.
This would be somewhat surprising since over a dozen stars have
already been analyzed in the relevant effective temperature range.
However, considering
the top panels of Figure~\ref{fig:lithium}, it is also remarkable that in
50 million years, stars between $4500$\,K and $5200$\,K go from
having a tight lithium sequence to one with a dispersion
$\approx$$10\times$ greater.  The existence of the Li dispersion in
Pleiades-age K-dwarfs has been known for decades; it has also been
known that the stars with the largest lithium abundances are also the
most rapidly rotating
\citep{butler_pleiades_1987,soderblom_evolution_1993}.  More recent
analyses of this correlation have been reviewed by
\citet{bouvier_lithium-rotation_2020}.  The conclusion of that work
was that the origin of the rotation-lithium correlation likely lies
within pre-main-sequence stellar physics.  If so, one would expect the
IC\,2602 and Tuc-Hor K-dwarfs to show a larger intrinsic lithium
dispersion.  A recent analysis of the $\approx$40\,Myr
NGC~2547 by \citet{binks_2022} suggests that this may
be the case, though that study only had $\approx$10 stars in the relevant
effective temperature range.
An alternative explanation could be that the overall metallicity of Cep-Her
is different from Tuc-Hor and IC\,2602, but this seems unlikely given the near-solar
metallicities we have measured for the Kepler Objects of Interest.
Broadly, these considerations suggest that Cep-Her is
a worthy object for further spectroscopic analyses of lithium near the
zero-age main sequence.

\subsubsection{H$\alpha$}

As shown in Figure~\ref{fig:koiyouthindicators}, H$\alpha$ is in
emission for both components of KOI-7913, and in absorption for the
hotter stars.  Additionally, the emission appears double-peaked for
both of the KOI-7913 components.  An important note is that KOI-7913~A and
KOI-7913~B were spatially resolved from each other during data
acquisition.  Performing a cross-correlation between each of the stars
and the nearest matches in the Keck/HIRES template library, we also
found that the CCFs for both components of KOI-7913 showed no
indications of double-lined binarity \citep{kolbl_detection_2015}.

Balmer line emission, particularly in H$\alpha$, is expected for
low-mass stars of this age.  \citet{kraus_stellar_2014} for instance, in their
survey of Tuc-Hor ($\approx$40\,Myr), observed that all
cluster members with spectral types $>$${\rm K4.5V}$ had H$\alpha$ in
emission.  This is consistent with our observations: KOI-7913 shows
H$\alpha$ in emission for both components, and in absorption
for all of our other Kepler objects (Figure~\ref{fig:koiyouthindicators}, lower panel).  The double-peaked nature of the
emission, though not always present, is also common for
active stars.  Proxima Centauri, for instance, has double-peaked
H$\alpha$ emission \citep{collins_calculations_2017}.  Given
that we have ruled out spectroscopic binarity, the most
likely
explanation is self-absorption: photons near the center of the line
see a greater optical depth from higher layers of the chromosphere,
while photons on the wings are too far from the
rest wavelength to excite electrons and be re-absorbed in the upper
layers.  The exact details of when a star's atmosphere reaches the
conditions for such self-absorption require non-local thermal
equilibrium models of the chromosphere
\citep{short_chromospheric_1998,2005A&A...439.1137F}.

\section{The Planets}
\label{sec:planets}

\begin{figure*}[tp]
	\begin{center}
		\leavevmode
		\subfloat{
			\includegraphics[width=0.85\textwidth]{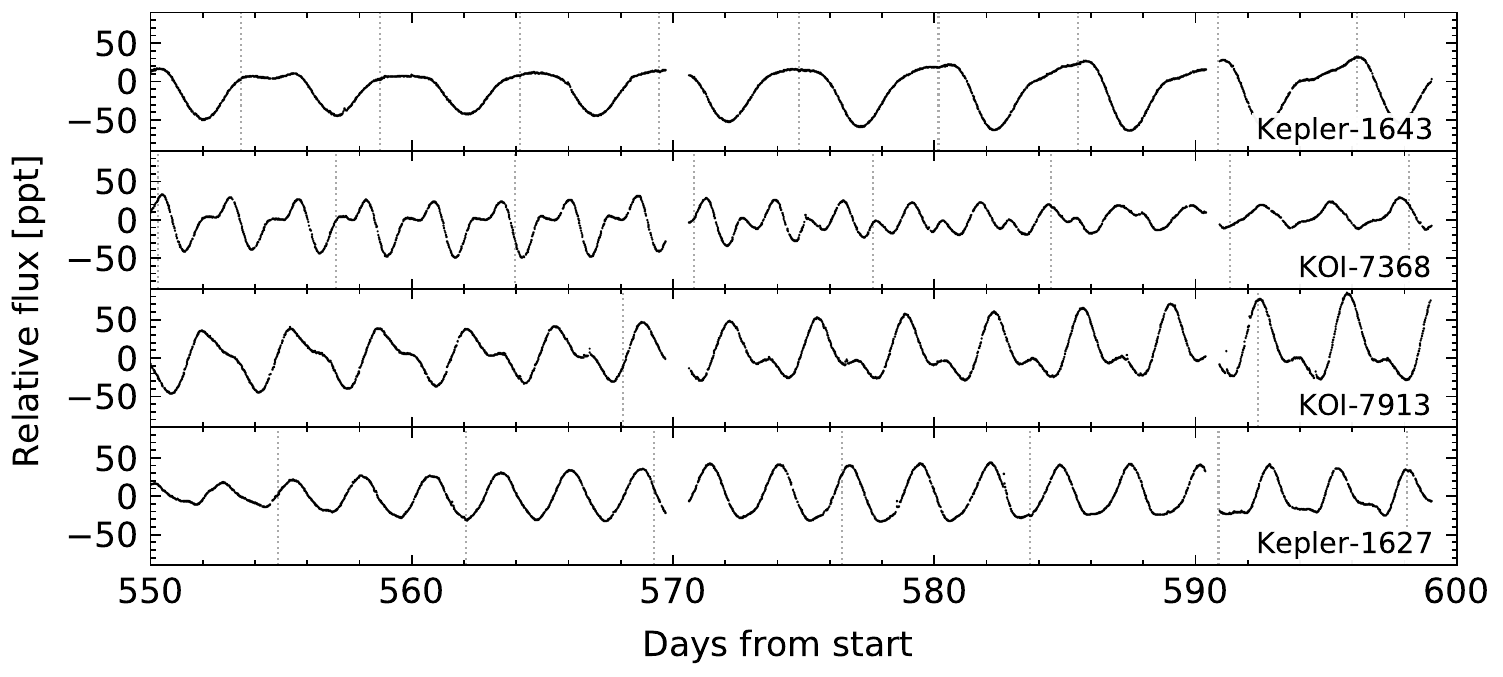}
		}

		\vspace{-0.5cm}
		\subfloat{
			\includegraphics[width=0.33\textwidth]{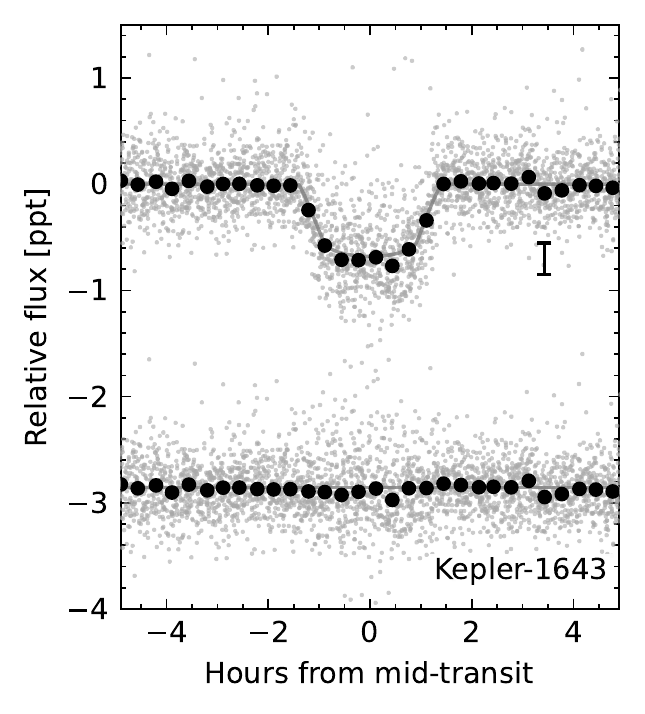}
			\includegraphics[width=0.33\textwidth]{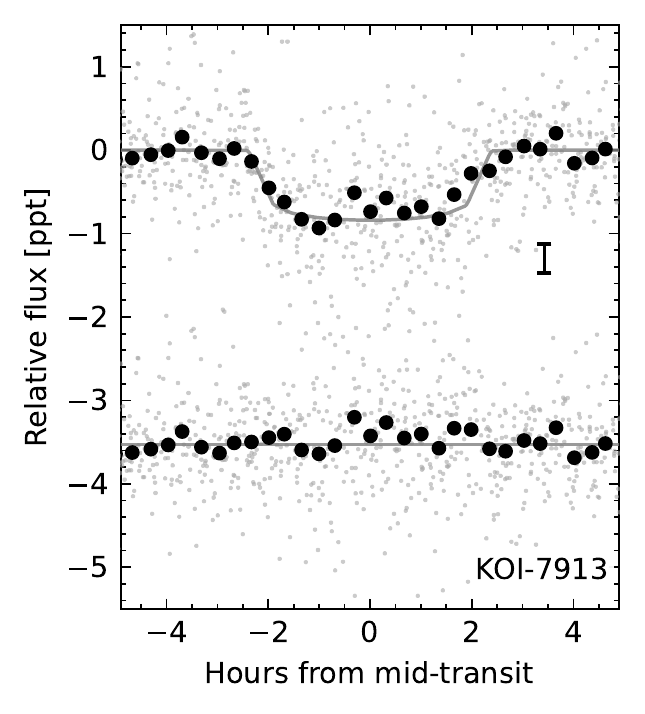}
		}

		\vspace{-1.27cm}	
		\subfloat{
			\includegraphics[width=0.33\textwidth]{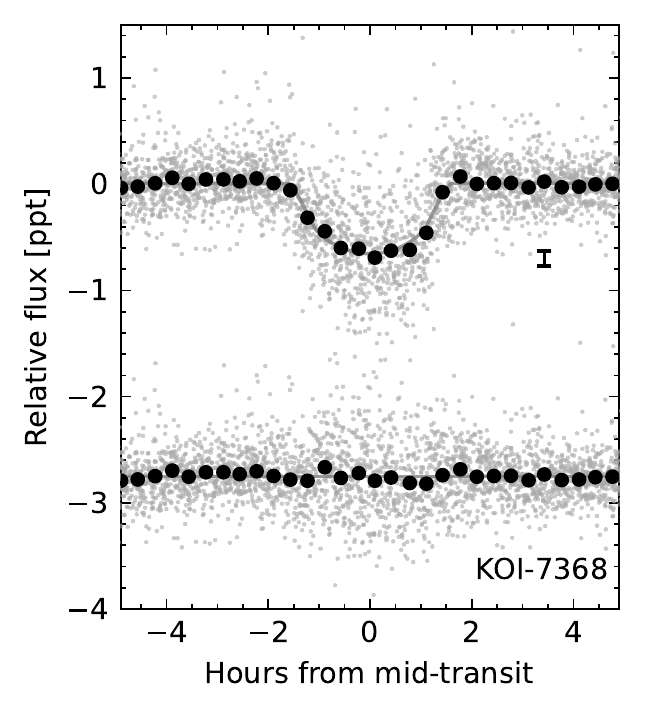}
			\includegraphics[width=0.33\textwidth]{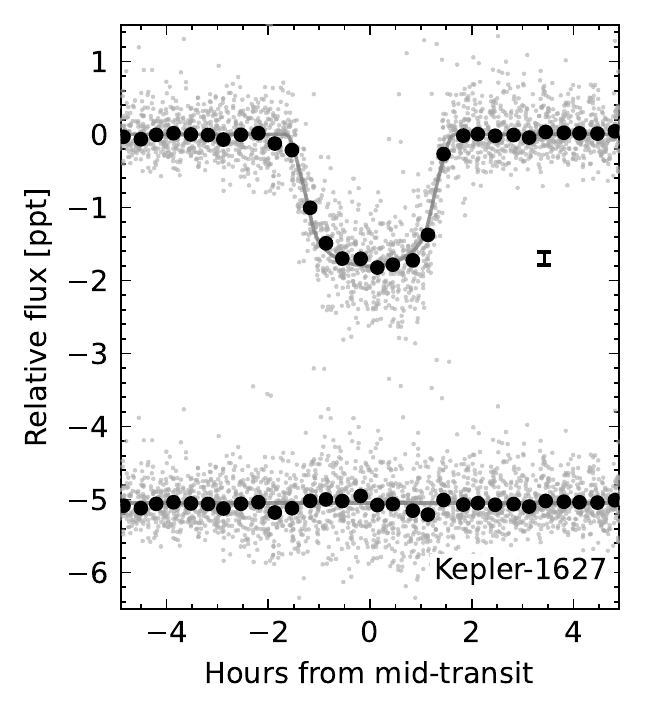}
		}
	\end{center}
	\vspace{-0.5cm}
	\caption{
		{\bf Raw and processed light curves for the Kepler Objects of
    Interest in Cep-Her.}   
    {\it Top}: 50 day light curve segment from the 3.9 years of Kepler
    data.  The ordinate shows the \texttt{PDCSAP} median-subtracted
    flux in units of parts-per-thousand ($\times 10^{-3}$).  The
    dominant signal is from starspots; planetary transit times
    are indicated with vertical dashed lines, but the individual transits are not visible at this scale.
    {\it Bottom}:
    Phase-folded transits of Kepler-1643, KOI-7913, KOI-7368, and
    Kepler-1627 with stellar variability removed.  The maximum a
    posteriori model is shown with the gray line, and the
    residual after subtracting the transit model is vertically
    displaced.  Windows over 10 hours are shown.  Gray points are
    individual flux measurements; black points are binned to 20 minute
    intervals, and have a representative 1-$\sigma$ error bar in the
    center-right of each panel. 
		\label{fig:planets}
	}
\end{figure*}

\subsection{Kepler Data}

The Kepler space telescope observed Kepler-1643, KOI-7913, and
KOI-7368 at a 30-minute cadence between May 2009 and April 2013.  For
all three systems quarters 1 through 17 were observed with minimal
data gaps.  The top panel of Figure~\ref{fig:planets} shows a 50-day
slice of the \texttt{PDCSAP} light curves for the three new Cep-Her
candidates, along with Kepler-1627.  In \texttt{PDCSAP},
non-astrophysical variability is removed through a cotrending approach
that uses a set of basis vectors derived by applying singular value
decomposition to a set of systematics-dominated light curves
\citep{smith_kepler_PDC_2017}.  In our analysis, we used the
\texttt{PDCSAP} light curves with the default optimal aperture
\citep{smith_finding_2016}.  Cadences with non-zero quality flags were
omitted.  In all cases, the stars are dominated by spot-induced
modulation with peak-to-peak variability between 2\% and 10\%.  These
signals are much larger than the transits, which have depth
$\approx$0.1\%.  To quantify the stellar rotation periods, we
calculated the Lomb-Scargle periodogram for each Kepler quarter
independently.  The resulting means and standard deviations are in
Table~\ref{tab:sysparams}.

\subsection{Transit and Stellar Variability Model}
\label{sec:fitting}

Our goals in fitting the Kepler light curves are twofold.  First, we
want to derive accurate planetary sizes and orbital properties.
Second, we want to remove the spot-induced variability signal to enable a
statistical assessment of the probability that the transit signals are
planetary.

We fitted the data as follows.  Given the transit
ephemeris from \citet{thompson_planetary_2018}, we first trimmed the
light curve to a local window around each transit that spanned three
transit durations before and after each transit midpoint.  The
out-of-transit points in each local window were then fitted with a
fourth-order polynomial, which was divided out from the light curve.
The resulting flattened transits were then fitted with a transit model
that assumed quadratic limb darkening.  The model therefore included 8
free parameters for the transit ($\{P, t_0, \log R_{\rm p}/R_\star, b,
u_1 ,u_2, R_\star, \log g\}$), 2 free parameters for the light curve
normalization and a white noise jitter ($\{\langle f \rangle, \sigma_f
\}$), and 5 fixed parameters for each transit.

We fitted the data using \texttt{exoplanet}
\citep{exoplanet:exoplanet}.  We assumed a Gaussian likelihood, and
sampled using \texttt{PyMC3}'s No-U-Turn Sampler
\citep{hoffman_no-u-turn_2014}, after having initialized to the the
maximum a posteriori (MAP) model.  We used the
\citet{gelman_inference_1992} statistic, $\hat{R}$, as our convergence
diagnostic.  The resulting fits are shown in the lower panels of
Figure~\ref{fig:planets}, and the important derived parameters are in
Table~\ref{tab:sysparams}.  The set of full parameters and their
priors are given in Appendix~\ref{app:transit}.

A potential drawback of our approach is that to remove
the starspot-induced variability, we fixed 5 
parameters per transit to their MAP values.
An alternative could be to fit the
planetary transits simultaneously with the starspot-induced
variability using a quasiperiodic Gaussian process (GP).  We explored
this approach, but ultimately prefer our model for its simplicity, and for the benefit that the white noise jitter never
trades off with any parameter equivalent to a damping timescale for
the coherence of the GP.  It is also computationally efficient, and it
captures the planetary parameters about which we care the most.

\subsection{Planet Validation}
\label{subsec:validation}

In the future, it may be possible to obtain independent evidence for
the planetary nature of the Cep-Her planets, for instance by observing
spectroscopic transits.  For now,
it is of interest whether the transit signals might be astrophysical
false positives, or whether they are statistically more likely to be
planetary.  We adopt the Bayesian framework implemented in
\texttt{VESPA} to assess the relevant probabilities
\citep{morton_efficient_2012,vespa_2015}.  Briefly summarized, the
priors in \texttt{VESPA} assume the binary star occurrence rate from
\citet{raghavan_survey_2010}, direction-specific star counts from
\citet{girardi_star_2005}, and planet occurrence rates as described by
\citet[][Section~3.4]{morton_efficient_2012}.  The likelihoods are
then evaluated by forward-modeling a synthetic population of eclipsing
bodies for each astrophysical model class, in which each population
member has a known trapezoidal eclipse depth, total duration, and
ingress duration.  These summary statistics are then compared against
the actual photometric data to evaluate the probabilities of false
positive scenarios such as foreground eclipsing binaries, hierarchical
eclipsing binaries, and background eclipsing binaries.

\paragraph{Kepler-1643}
Kepler-1643~b (KOI-6186.01) was already validated as a transiting
planet by \citet{morton_false_2016}, who found a probability for any
of the aforementioned false positive scenarios of 9$\times$10$^{-6}$.
Repeating the calculation with our own stellar-variability correction
and the new NIRC2 imaging constraints, we find ${\rm FPP} =
6\times10^{-9}$.  Figure~\ref{fig:planets} shows the justification:
the transit is flat and has a high S/N ($\approx$$47$).  The shape is
therefore nearly impossible to reproduce with eclipsing binary models.

Intriguingly, Kepler-1643 failed one of the data validation
centroid shift tests (see the \texttt{q1\_q17\_dr25\_koi} data
release): the angular distance between the target star's KIC catalog
position and the position of the transiting source was measured as
$1\farcs0$ at 4.4-$\sigma$.  The reports show however that two
outlying quarters (2 and 6) drive the offset --- the centroid locations
from the other Kepler quarters are consistent at $\lesssim 0\farcs4$ (3-$\sigma$).
\citet{bryson_2013} showed that for typical field star KOIs without
centroid offsets, the mean offset distribution peaks at 0.3$''$ (their
Figure 23).  By comparison, stars with centroid offsets that can be
localized to nearby stars have a distribution that peaks at 7$''$ (their
Figure 32).  The stellar
variability in Kepler-1643 complicates the centroid-based
vetting tests, because the shifts measured by these tests are
determined from the in- and out-of-transit flux-weighted centroids.  For stars
with significant spot-induced variability there is no static baseline in either
the in- or out-of-transit phases, and so the centroid location may shift
depending on the rotation phase combined with the local scene.  Based on
these considerations, the centroid-level diagnostics for Kepler-1643 appear to
be consistent with the transit signal being localized to the target star.

\paragraph{KOI-7368}
KOI-7368.01 is listed on the NASA Exoplanet Archive as a ``candidate''
planet.  \citet{morton_false_2016} did not compute a false positive
probability for the system because their default trapezoidal fitting
routine failed, presumably due to the spot-induced variability.  Our
fitting approach rectifies this point, and our new NIRC2 images
revealed no new stellar companions.  Performing the relevant
calculation, we find ${\rm FPP} = 4\times10^{-3}$.  Though not
as convincing as Kepler-1643, this clears the threshold probability of 1 in 100 suggested by \citet{morton_false_2016} for calling a planet statistically validated.  The S/N of the transit is
$\approx$$32$, which indicates that it is unlikely to be caused by
systematic noise in the light curve (see Figure~\ref{fig:planets}).
The positional probability\footnote{Columns $\textit{pp\_host\_rel\_prob}$ and $\textit{pp\_host\_prob\_score}$ on the KOI Positional Probabilities table at the NASA Exoplanet Archive \citep{Akeson13}.} calculated by \citet{2017ksci.rept...16B} also indicates that the transit signal shares its position with the target star.

It bears mentioning that KOI-7368 shows a centroid shift in the
\texttt{q1\_q17\_dr25\_koi} validation reports, similar to
Kepler-1643.  For KOI-7368, the reported offset is smaller, and less
formally significant ($0\farcs2$; 3.0-$\sigma$).  Again, the
data validation reports show that the shift is caused by a few
outlying quarters (4, 5, 8, and 12).  Since the remaining
quarters show
consistent scatter in their centroid locations, these outlying
quarters are likely also caused by the stellar variability,
because their directions are inconsistent across different quarters.
Our NIRC2 imaging independently shows that there are no known
neighboring sources that could cause an offset of the observed
amplitude, as is also the case for Kepler-1643.

\paragraph{KOI-7913}
KOI-7913.01 is also currently listed on the NASA Exoplanet Archive as a
``candidate'' planet.  The \citet{morton_false_2016}
analysis was of Q1-Q17 KOIs from DR24, and therefore spanned KOI-1.01
to KOI-7620.01 (omitting KOI-7913.01).  However the results of the
subsequent DR25 analysis by Morton et al.\ are listed at the NASA
Exoplanet Archive.  The relevant table gives a probability for the
system being an astrophysical false positive of $1.4\times10^{-4}$,
with the most likely false positive scenario being a blended eclipsing
binary.  Repeating the calculation with our new detrending and
NIRC2 contrast curves, we find a similar result: ${\rm FPP} =
1.3\times10^{-4}$.  Though the transit has the lowest S/N of any of
the objects discussed ($\approx$$14$), its low FPP  can be understood through its flat-bottomed shape,
combined with its long transit duration relative to most
eclipsing binary models (Figure~\ref{fig:planets}).  The positional probability calculation performed by \citet{2017ksci.rept...16B} yielded a near-unity probability that the transit event is at the same location as the host star, and so the cumulative evidence suggests that KOI-7913~Ab is indeed a statistically validated planet.  Its
disposition has however previously fluctuated from ``false positive''
to ``candidate'' (see Appendix~\ref{app:koi7913}).  The most likely
explanation is the presence of KOI-7913~B, which is located
$\approx0.9$ Kepler pixels away from Kepler-7913 A.  
While the $\approx$1.5 pixel FWHM of the Kepler pixel response function
implies that there is blending between the two stars,
the target-pixel level data for KOI-7913 B reveals an
entirely different stellar rotation
period (Table~\ref{tab:sysparams}), and no hint of the transit signal.
This implies that KOI-7913 B cannot host the planet.

\section{Discussion \& Conclusion}
\label{sec:disc_conc}

\begin{figure*}[!t]
	\begin{center}
		\leavevmode
		\includegraphics[width=0.93\textwidth]{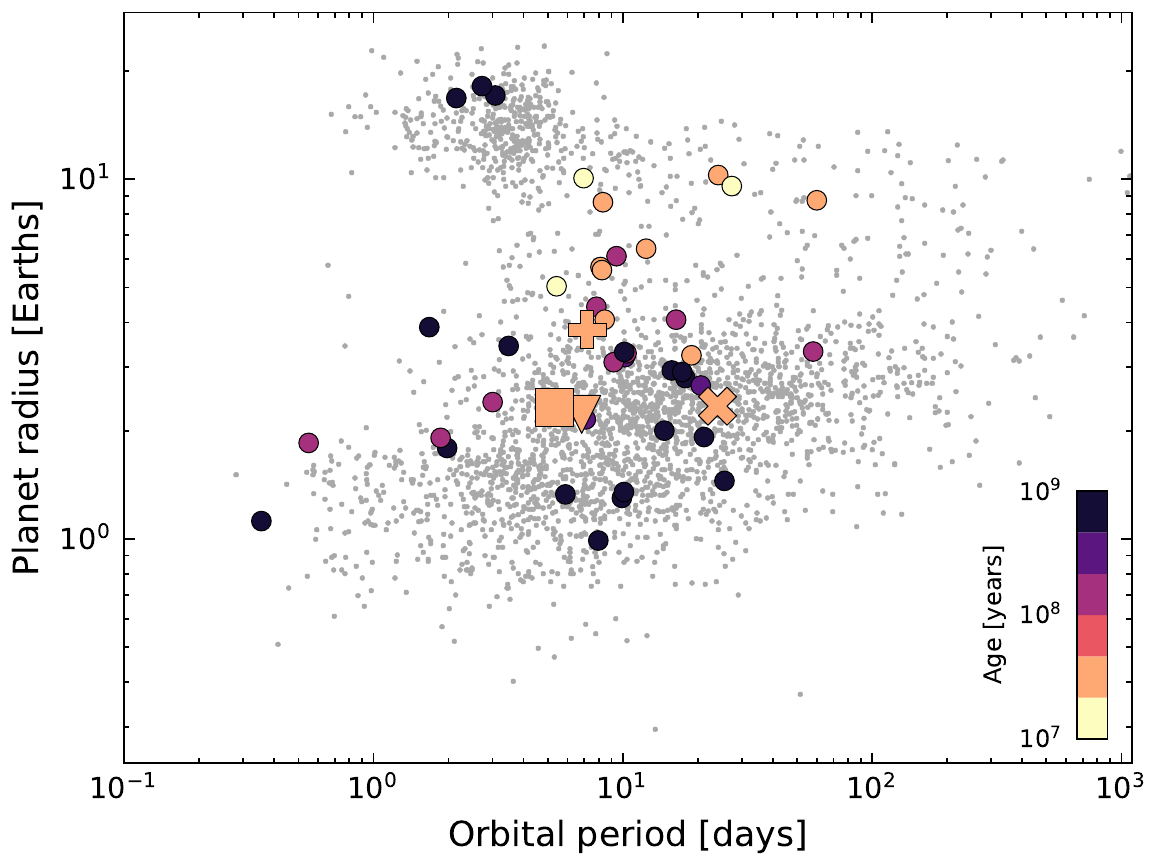}
	\end{center}
	\vspace{-0.6cm}
	\caption{
		{\bf Radii, orbital periods, and ages of transiting exoplanets}.
    Planets younger than a gigayear with ages more precise than a
    factor of three are emphasized. The Cep-Her planets are
    Kepler-1643~b ($\square$), KOI-7368~b ($\triangledown$),
    KOI-7913~Ab (X), and Kepler-1627~Ab (+).  Interesting trends in
    the population of planets younger than $10^8$ years old include {\it i)} their
    large sizes and {\it ii)} the lack of hot Jupiters.  The new
    objects of interest in Cep-Her have normal mini-Neptune sizes
    between 2 and 3\,$R_\oplus$, which is a novelty given their ages.
    Parameters are from the \citet{PSCompPars}.
		\label{fig:rp_period_age}
	}
\end{figure*}

\subsection{Normal-Sized Mini-Neptunes Exist at 40$\,$Myr}
\label{subsec:sizes}

The most significant novelty about the planets in Kepler-1643,
KOI-7368, and KOI-7913 is that their sizes (2.2 to 2.3\,$R_\oplus$)
are normal relative to the known population of mini-Neptunes from
Kepler.  At field star ages, mini-Neptune sizes span 1.8\,$R_\oplus$
to 3.6\,$R_\oplus$, with the most common size being $\approx
2.4\,R_\oplus$
\citep{Fulton_et_al_2017}.  The known planets younger than
$10^8$ years are almost all larger, with sizes between 4 and
10$\,R_\oplus$
\citep{Mann_K2_33b_2016,David_et_al_2016,benatti_possibly_2019,david_four_2019,newton_tess_2019,rizzuto_tess_2020,bouma_cluster_2020,mann_toi1227_2022}.
Figure~\ref{fig:rp_period_age} explores this by showing
the sizes, orbital periods, and ages of the known transiting planets,
emphasizing planets with precise ages.  The smallest previously known
planets comparable to the new Cep-Her mini-Neptunes are AU~Mic~c
($3.0\pm0.2\,R_\oplus$, see \citealt{martioli_aumicbc_2021} and
\citealt{gilbert_flares_2022}), Kepler-1627~Ab ($3.8\pm0.2\,R_\oplus$;
\citealt{bouma_kep1627_2022}), and AU~Mic~d ($4.2\pm0.2\,R_\oplus$;
\citealt{plavchan_planet_2020}).

The theoretical expectation is that mini-Neptunes with sizes of 2 to
3\,$R_\oplus$ should be common at ages of $10^7$ to $10^8$ years.
This expectation is tied to inferences about the initial distributions
of planetary core mass, core composition, and atmospheric mass
fraction \citep{owen_evaporation_2017}.  The Kelvin-Helmholtz cooling
timescale, which is tied to the entropy of the planetary interior
shortly after disk dispersal, also plays a significant role
\citep{owen_constraining_2020}.  As an example,
\citet{rogers_unveiling_2021} predicted that given a core mass
distribution peaked at $\approx$$4$\,$M_\oplus$, an ice-poor rock/iron
core composition, and a typical H/He mass fraction of $\approx$4\%,
there should be a single local maximum in planet occurrence rates at 2 to 3\,$R_\oplus$,
at times between 10 and 100\,Myr. In other words,
\citet{rogers_unveiling_2021} predict the existence of a ``radius
mountain'' at these early times, rather than a ``radius valley''.  The models advanced by
\citet{gupta_signatures_2020} and \citet{lee_primordial_2021}
agree that this local maximum should exist;
their differences lie in the mechanism for producing the radius
valley, and in whether a population of rocky planets is predicted to exist
at the time of disk dispersal. 

Systems such as K2-25, V1298~Tau, HIP-67522, TOI-837, and TOI-1227
have sizes that are anomalously large relative to the predicted peak
in planet occurrence at 2 to 3\,$R_\oplus$.  However, their large
sizes can be accommodated by invoking any of {\it i)} larger core
masses, {\it ii)} more volatile-rich compositions, {\it iii)}  larger
initial atmospheric mass fractions, or {\it iv)} longer thermal
cooling times.  Secure mass measurements would help constrain this
parameter space, but the $\sim$1\,\kms\ spot-induced radial velocity
semi-amplitudes make measuring the Doppler orbits very difficult
\citep[][]{cale_diving_2021,zicher_one_2022,klein_one_2022}.
Regardless, the new Kepler-1643, KOI-7368, and KOI-7913 systems do
demonstrate that at least some planets at 40\,Myr have sizes that are
consistent with theoretical expectations for mini-Neptunes.  While
selection effects imposed by spot-induced photometric variability are
a likely explanation for why planets this small have not previously
been identified \citep[{e.g.},][]{zhou_2021_tois}, future work should
quantify this bias more carefully, in order to enable empirical
studies of how the planetary size distribution changes at early times.

\subsection{Is CH-2 a Coeval Population?}
\label{subsec:ch2}

RSG-5, and Kepler-1643's membership inside it, meet typical
expectations for a star claimed to be in an open cluster.  RSG-5 is an
obvious overdensity relative to the local field, and our membership
selection easily yielded a clean pre-main-sequence locus
(Figure~\ref{fig:age}).  CH-2, and KOI-7913 and KOI-7368's membership
inside it, do not meet these expectations in as obvious a manner.
This is because the CH-2 association is diffuse.

To quantify the density difference between CH-2 and RSG-5, we can
compare the spatial and velocity volumes searched for each group.  For
RSG-5, we drew \nrsgfive\ candidate members from a $30\,{\rm
pc}\times30\,{\rm pc}\times40\,{\rm pc}$ rectangular prism, given a
$1.5 {\rm \mkms} \times 2.5 {\rm \mkms }$ rectangle in apparent
galactic velocity.  For CH-2, our \nchtwo\ candidate members came from
a rectangular prism of dimension $50\,{\rm pc}\times40\,{\rm
pc}\times30\,{\rm pc}$, and a rectangular box of $2 {\rm \mkms} \times
4 {\rm \mkms}$.  If we define the searched volume in units of ${\rm
pc}^3\,{\rm km^2}\,{\rm s^{-2}}$, then the volume ratio of CH-2 to
RSG-5 is 3.5 to 1.  The ratio of number densities (candidate members
per unit searched volume) in RSG-5 relative to CH-2 is 16 to 1.

Given its low density, is CH-2 truly a star cluster?  For this
discussion, we adopt the definition that a star cluster is a group of
at least 12 stars that was physically associated at its time of
formation.  The ``12'' is set to distinguish star clusters from
high-order multiples \citep[see][]{krumholz_star_2019}.  We explicitly
do not require a ``star cluster'' to be gravitationally bound:
dissolved clusters as well as their tidal tails are included in our
adopted definition of ``clusters''.  We similarly do not require a
threshold number of stars per unit spatial volume.  The latter point
acknowledges that an important factor in cluster identification is
also coherence in velocity space.  For instance, the Psc-Eri stream, which has a shape that can be approximated as a 600 parsec-long cylinder with a radius of 30 parsecs, has a number density roughly a factor of three times lower than even CH-2 \citep{roser_psceri_2020}.  However its existence is discernible because of the $\lesssim 2.5$\kms\ scatter in its cylindrical velocities. Perhaps once stellar rotation periods
and chemical abundances reach the same level of ubiquity as stellar
proper motions, they might enable further refinement in our ability to
discover stars that formed as part of the same event.

From a data-driven perspective, demonstrating that a group of stars
was physically associated at its time of formation is challenging.
While some young groups show kinematic evidence for expansion
\citep{kuhn_kinematics_2019}, many, including Sco-Cen, do not
\citep{wright_kinematics_2018}.  This complicates the feasibility of
deriving kinematic ages through traceback, as well as through the
expansion itself \citep[see][]{crundall_chronostar_2019}.  A more
minimal approach is that suggested by \citet{tofflemire_2021}: search
for coeval, phase-space neighbors, measure their ages, and determine
if they share a common age.  This approach can demonstrate whether a
star is currently associated with a set of coeval stars, though it
falls short of determining what the association looked like in the
past.  Our analysis of CH-2 meets the latter standard for
demonstrating the existence of a $\approx$40\,Myr stellar association. 

It would be a worthy exercise to perform a similar
search for coeval phase-space neighbors on the entire dataset of known
exoplanet hosts.  For the time being, we can offer the anecdotal point
that in our experience, most stars do not have dozens of 40\,Myr
neighbors within a local volume of a few \!\kms\ and tens of parsecs.

\subsection{Future work}

\paragraph{Cep-Her}
Our analysis to date has focused only on portions of Cep-Her that were
observed by Kepler: RSG-5, CH-2, and the $\delta$~Lyr cluster.  In
\citet{bouma_kep1627_2022} as well as this work, we have shown that
these groups share similar ages, and have kinematic correlations that
suggest a common origin.  With that said, the membership and
kinematics of the other Cep-Her groups shown in
Figure~\ref{fig:XYZvtang} deserve independent attention.  An important
aspect of the remaining work will be to acquire radial velocities for a larger subset of the stars, and to determine whether the traceback approach
could be applicable.  Wide-field spectroscopic surveys such as LAMOST
\citep{zhao_2012_LAMOST} or SDSS-V \citep{kollmeier_2017} could enable
such analyses for the brightest members, while also providing sensitivity to the Li 6708\,\AA\
line.  The Gaia DR3 RVS spectra (released during review of this manuscript) could contain similar velocity information down to spectral types of $\approx$K5V ($G_{\rm RVS}\lesssim 14$), and perhaps also enable analyses of the calcium infrared triplet as a youth indicator.  The combination of more complete kinematics and youth indicators would help in definitively unraveling the formation
history of the complex.

A number of worthy photometric projects also seem possible given the
new understanding of Cep-Her.  One is asteroseismology of the
$\delta$~Sct stars, using either TESS or Kepler data
\citep{bedding_very_2020}.  For cases in which the modes are resolved,
this might yield age or metallicity estimates for the subgroups
independent of other methods.  Other projects could include a more
comprehensive analysis of the stellar rotation periods,
searches of the Kepler light curves for exocomets
\citep{zieba_transiting_2019}, and searches for missed planets around
the most rapid rotators.

\paragraph{Exoplanet demographics at early times}
Our main motivation for finding new young planets is to help benchmark
models for planetary evolution.  However demographic analyses of the
known planets between $10^7$ and $10^9$ years have so far been rather
limited.  Approximately 40 such planets are now known
(Figure~\ref{fig:age}).  About half come from K2, a quarter from TESS,
and now a quarter from Kepler.

Given the current state of the field, a few reflections regarding
experimental design of a demographic survey focused on planetary
evolution over the first gigayear might be useful.  The first is that
such a project requires a set of target stars with known ages.  A
promising way to compile relevant stars could be to combine automated
spatio-kinematic clustering from Gaia with rotation periods measured
using TESS \citep[see the appendices of][]{bouma_kep1627_2022}.  The
second consideration is that all the known young planets smaller than
$3$\,$R_\oplus$ come from either K2 or Kepler.  Demographic inferences
based on TESS are therefore limited to planetary sizes
$\gtrsim4$\,$R_\oplus$, for planets close-in to their host stars.  It
would be worthwhile to compare the occurrence rates of both types of
planets with those from the main Kepler sample.  One specific question
that seems within reach would be to clarify whether enough young stars
have been searched for the dearth of young hot Jupiters to be
significant.  Since the hot Jupiter occurrence rate is strongly
dependent on stellar mass and metallicity
\citep{petigura_metallicity_2018,petigura_cksX_2022}, particular care
would be needed to select a sample of well-studied FGK dwarfs for the
measurement, likely using stars in Sco OB2, Cep-Her, and Orion.  For
demographic studies focused on how mini-Neptune sizes evolve, the
combined K2 and Kepler dataset would be the better primary source.

\subsection{Summary}

We have shown that Kepler-1643~b, KOI-7368~b, and KOI-7913~Ab are 40
to 50 million years old, and that each system is most likely
planetary.  The evidence for the planetary interpretation comes from
an application of \texttt{VESPA} to the Kepler data, alongside new
imaging from NIRC2.  The validity of the \texttt{VESPA} framework
rests on the premise that non-astrophysical false positives can be
rejected.  
This seems to be the case for all three objects, even though
Kepler-1643 and KOI-7368 
both show weak centroid offsets in specific quarters.  For both systems, the observed shifts
are consistent with being caused by starspot-induced variability in
specific quarters spuriously moving the stellar center-of-light.
Independently, our imaging rules out companion stars with the
brightnesses and positions that would be needed to explain the
reported shifts.  All three objects are therefore most likely planets.

Each system has multiple indicators of youth that support the reported
ages.  For Kepler-1643, the strongest youth indicator is its physical
and kinematic association with RSG-5.  Based on the color--absolute
magnitude diagram, we are able to select members of this cluster with
a false positive rate of a few percent (Figure~\ref{fig:age}).
Kepler-1643 is one such member.  While the stellar rotation period
period agrees with this assessment, the star's lithium equivalent
width is marginally low, which might motivate future exploration of
lithium depletion across FGKM stars in RSG-5 (see
Section~\ref{subsec:specyouth}).

The spatio-kinematic argument for the youth of KOI-7368 and KOI-7913
is weaker because they are in an association of stars, CH-2, that is
more diffuse.  For KOI-7913, stronger indicators of its age come from
its binarity.  Both stellar components in KOI-7913 have isochronal ages
consistent with $40$\,Myr.  Both components also show H$\alpha$ in
emission, which for the transit-hosting $\approx$K6V primary is a strong indicator
that the star is $\lesssim$100$\,$Myr old.  KOI-7368 is  more
massive, and its Li 6708\,\AA\ measurement and stellar rotation period
provide independent verification of the star's youth.

The astrophysical implication of these considerations is that planets
$\approx$2 Earth radii in size exist at ages of 40 million years.  It
will be interesting to continue the push down to smaller planetary
sizes at comparable ages -- the planetary detections we have presented
are well above the average detection significance for Kepler planets.
There may still be room at the bottom.


\acknowledgements
L.G.B{.} is supported by the Heising-Simons Foundation 51
Pegasi~b Fellowship and the NASA TESS GI Program (80NSSC21K0335
and 80NSSC22K0298).
R.K{.} is supported by the Heising-Simons Foundation.
J.L.C{.} is supported by NSF AST-2009840 and the NASA TESS GI Program
(80NSSC22K0299).
D.H. is supported by the Alfred P. Sloan Foundation and NASA (80NSSC19K0597).

\software{
  \texttt{astropy} \citep{astropy_2018},
  \texttt{astroquery} \citep{astroquery_2018},
  \texttt{exoplanet} \citep{exoplanet:exoplanet}, and its
  dependencies \citep{exoplanet:agol20, exoplanet:kipping13, exoplanet:luger18,
   	exoplanet:theano},
  \texttt{PyMC3} \citep{salvatier_2016_PyMC3},
  \texttt{tesscut} \citep{brasseur_astrocut_2019},
  \texttt{unpopular} \citep{hattorio_2021_cpm},
\texttt{VESPA} \citep{morton_efficient_2012,vespa_2015},
}
\ 

\facilities{
 	{\it Astrometry}:
 	Gaia. 
 	{\it Imaging}:
    Second Generation Digitized Sky Survey. 
 	Keck:II~(NIRC2).
 	{\it Spectroscopy}:
	Tillinghast:1.5m~(TRES).
 	Keck:I~(HIRES).
 	{\it Photometry}:
	  Kepler,
 	  TESS,
    ZTF.
}

\clearpage
\bibliographystyle{yahapj}                            
\bibliography{bibliography} 

\appendix
\section{Candidate Cep-Her Members}
\label{app:members}

\paragraph{Table~2} contains 338 candidate Cep-Her members with weights
$D>0.02$ observed by Kepler.  The complete catalog of candidate
Cep-Her members will be provided by R.~Kerr et al.\ in prep.\ using Gaia DR3; Table~2 is
from an early version of that analysis based on Gaia EDR3.  Note that more
restrictive weight cuts should be imposed if one wishes to remove
the majority of field star interlopers.
Table~2 was created by cross-matching candidate Cep-Her members
(selected using Gaia EDR3; Section~\ref{subsec:members}) against a
Kepler to Gaia DR2 cross-match (the \texttt{gaia-kepler.fun}
crossmatch database created by Megan Bedell).  The
\texttt{kic\_dr2\_ang\_dist} column is from the latter table.  The
EDR3 to DR2 match was performed using the
\texttt{gaiaedr3.dr2\_neighbourhood} table, and the closest proper
motion and epoch-corrected angular distance neighbor was taken as the
single best match.  The \texttt{edr3\_dr2\_mag\_diff} column gives
some indication of the reliability of this EDR3 to DR2 conversion, as
there are a few cases between Gaia DR2 and EDR3 where partially
resolved binaries became fully resolved.

\paragraph{Candidate matches between Cep-Her and the Kepler Objects of
Interest:}
The full list of candidate matches between Cep-Her and the Kepler
Objects of Interest is as follows -- the objects are listed in order
of descending weights, $D$.
Objects designated as confirmed planets included
Kepler-1627,
Kepler-1643,
Kepler-1331,
Kepler-1062, and
Kepler-1933.  
Objects designated as candidate planets included
KOI-5264,
KOI-8007,
KOI-7572,
KOI-7375,
KOI-7368,
KOI-7638,
KOI-5632, and
KOI-7913.
Objects designated known false positive planet candidates included
KOI-6437, 
KOI-5988, 
KOI-7871,
KOI-7655,
KOI-5024,
KOI-61,
KOI-4336,
KOI-6812,
KOI-3399, and
KOI-6277.
Finally, Kepler-1902 (KOI-3090) has one confirmed planet
(KOI-3090.02), and one false positive (KOI-3090.01).
Of these objects, only Kepler-1627, Kepler-1643, KOI-7368, and
KOI-7913 met our requirements for potentially
both {\it i)} having real planets, and {\it ii)} being $\lesssim 10^8$
years old, based on the presence of rotational modulation at the
expected period and amplitude.  
Of the
14 confirmed and candidate planets,
6 failed the first filter, and 7 independently failed the second.
One object was ambiguous: Kepler-1933.
This system has a confirmed $\approx$$1.4\,R_\oplus$ planet, a stellar
rotation period of 6.5\ days, and an effective temperature of $\approx$$5750\,{\rm K}$.  This places it near the upper envelope of the
rotation period vs.\ color distribution for the Pleiades, making
it unlikely to be $\approx$40\,Myr old.  Nonetheless, we acquired a
reconnaissance HIRES spectrum, and it yielded ${\rm EW}_{\rm Li} = 93 \pm
5$\,m\AA.  Combined with the rotation period, this suggests an age
for Kepler-1933 between 100 and 300\,Myr.  Based on these indicators, the system is
unlikely to be part of Cep-Her, but could merit further study.

\paragraph{Table~3} contains spatial, kinematic, astrometric, and
rotation period information for the \nrsgfive\ candidate RSG-5 members
and \nchtwo\ candidate CH-2 members described in
Section~\ref{subsec:members}.  These are the data used to make the
lower panels of Figure~\ref{fig:age}; as with Table~2, these are
from a preliminary version of the SPYGLASS 1\,kpc expansion (R. Kerr et
al.\ in prep). 
We adopted the ZTF period over the
TESS period in three cases: (1) Gaia EDR3 2081755809272821248: the top ZTF Lomb-Scargle peak gave 6.61 days, while our default pipeline favored a TESS peak of 13.34 days;  manual inspection of the light curve favors the former; (2) Gaia EDR3 2081737529891330560: we found 3.06
days with TESS and 6.64 days with ZTF; we suspect that TESS captured
the 1/2-period harmonic and adopt the approximately double value from
ZTF; (3) 2134851775526125696: for this star, we measured 1.91 days
with TESS from Cycle 2, but noted that the signal appeared to be
missing in Cycle 4; ZTF found a strong signal at 12.23 days and we
adopt this as the star's period. In the remaining overlap cases, we
adopted the average between TESS and ZTF as the final period. For
these overlap stars, the median absolute deviation is 0.01 days,
showing remarkable consistency between the surveys. For three stars,
we failed to detect a period in TESS but recovered one from ZTF; in
all cases the periods appear to be 13--16 days.  These stars were: (1)
Gaia EDR3 2129930258400157440, for which TESS showed a flat light
curve while ZTF yielded a 15.3-day period; (2) Gaia EDR3
2082376861542398336, LS found a 7.6-day period which we rejected
during visual validation; we found 15.4 days with ZTF, and we suspect
that the weak/rejected signal form TESS might have been a 1/2 period
harmonic; (3) Gaia EDR3 2082397099429013120, similar to the previous
case, we rejected a 6.7-day signal from TESS and recovered a 12.8-day
period with ZTF.

\section{Table of Transit Fit Parameters}
\label{app:transit}

Table~4 gives the full set of fitted and derived
parameters from the model described in Section~\ref{sec:fitting}.
Priors and convergence statistics are also listed.

\section{Disposition History of KOI-7913}
\label{app:koi7913}

The disposition of KOI-7913.01 has been debated: in
\texttt{q1\_q17\_dr25\_koi} the source was flagged as a false
positive, with the comment ``cent\_kic\_pos---halo\_ghost''.  This
comment and disposition were removed in the
\texttt{q1\_q17\_dr25\_sup\_koi} data release, which renamed the
planet a ``candidate''.  In this note, we discuss the interpretation
of these flags (which do not apply to the system, according to the
latest analysis).  We also discuss how the relative on-sky positions of
KOI-7913~A and KOI-7913~B affect the interpretation of the Kepler
data.

As described by \citet{thompson_planetary_2018}, the
``cent\_kic\_pos'' flag is an indication that the measured source
centroid is offset from its expected location in the Kepler Input
Catalog.  The final Kepler data validation reports, generated 2016 Jan
30, do not show this to be the case for KOI-7913.  Moreover, the
statistical significance of any centroid offset is lower than for
KOI-7368 and Kepler-1643 (which both show centroid offsets that are
likely explained by the stellar variability).

What of the ``halo\_ghost'' flag?  This test measures the transit
strength for the pixels inside the aperture, and compares it to that
measured in the ring of pixels around said aperture (the ``halo'').
One usually expects the transit signal to be strongest in the
central aperture, rather than the halo.  Two types of false positive
scenarios can change this and trigger the flag: the first is when
optical ghosts from bright eclipsing binaries reflect off the CCD, and
contaminate the target star.  The second is when the PRF of nearby
stars directly overlaps with the PRF of the target star (see
\citealt{thompson_planetary_2018}, Section A.5.2).  The most obvious
explanation for KOI-7913 is the latter case, given that KOI-7913 B is
$\approx0.9$ Kepler pixels away from Kepler-7913 A and so it usually
part of the ``halo''.  Due to the on-sky orientation of KOI-7913 A and
KOI-7913 B, the default ``optimal aperture'' selected in quarters 3,
7, 11, and 15 in fact included both stars, while for the remaining
quarters KOI-7913 B was excluded from the optimal aperture but was
included as part of the halo (see pages 35 through 71 of the data
validation reports.)  

Given the orientation of the stars and the $\approx$1.5 pixel FWHM of
the Kepler pixel response function, some blending between the two
stars is present.  The pointing geometries from quarters 3, 7, 11, and
15 however did not affect the observed transit depths,
which is an indication that the crowding metric applied in the data
products accurately correct the mean flux level
\citep{2017ksci.rept....6M}.  Analysis of the target-pixel data that
was separately acquired for KOI-7913~B also reveals a
different stellar rotation period, and no hint of the transit signal.

\begin{deluxetable}{lll}
    

\tabletypesize{\footnotesize}


\tablecaption{Candidate Cep-Her members observed by Kepler}
\label{tab:cepherkepler}
\tablenum{2}

\tablehead{
  \colhead{Parameter} &
  \colhead{Example Value} &
  \colhead{Description}
}
\startdata
\texttt{dr2\_source\_id}      & 2073765172933035008 & Gaia DR2 source identifier. \\
\texttt{dr3\_source\_id}      & 2073765172933035008 & Gaia (E)DR3 source identifier. \\
\texttt{kepid} & 5641711      & KIC identifier. \\
\texttt{ra} &    297.40986     & Gaia EDR3 right ascension [deg]. \\
\texttt{dec} &   40.89719      & Gaia EDR3 declination [deg]. \\
\texttt{weight} & 0.041    & Strength of connectivity to other candidate cluster members. \\
\texttt{v\_l} & -0.51     & Longitudinal galactic velocity, $v_{l^*}$ [km\,s$^{-1}$]. \\
\texttt{v\_b} & -8.23     & Latitudinal galactic velocity [km\,s$^{-1}$]. \\
\texttt{x\_pc} & -8035.4  & Galactocentric $X$ position coordinate [pc]. \\
\texttt{y\_pc} & 331.4     & Galactocentric $Y$ position coordinate [pc]. \\
\texttt{z\_pc} & 65.3      & Galactocentric $Z$ position coordinate [pc]. \\
\texttt{kic\_dr2\_ang\_dist} & 0.298 & Separation between KIC and Gaia DR2 positions [arcsec]. \\
\texttt{edr3\_dr2\_mag\_diff} & 0.002 & $G$-band difference between EDR3 and DR2 source match [mag]. \\
\enddata
\tablecomments{Table~2 is published in its entirety in a machine-readable
format.  One entry is shown for guidance regarding form and content.  \added{Users who wish to minimize field star contamination should apply more restrictive weight cuts, {\it e.g.}, $\texttt{weight}>0.1$.}}
\vspace{-0.5cm}
\end{deluxetable}

\vspace{-6cm}
\begin{deluxetable}{lll}
    

\tabletypesize{\footnotesize}


\tablecaption{Rotation periods and kinematics for candidate RSG-5 and
  CH-2 members.}
\label{tab:rot}
\tablenum{3}

\tablehead{
  \colhead{Parameter} &
  \colhead{Example Value} &
  \colhead{Description}
}
%
\startdata
\texttt{dr3\_source\_id}      & 2127562009133684480 & Gaia (E)DR3 source identifier. \\
\texttt{ra} &    291.02306     & Gaia EDR3 right ascension [deg]. \\
\texttt{dec} &  46.43843      & Gaia EDR3 declination [deg]. \\
\texttt{parallax} &  3.7099      & Gaia EDR3 parallax [milliarcsec]. \\
\texttt{ruwe} &  0.981      & Gaia EDR3 renormalized unit weight error. \\
\texttt{weight} & 0.087    & Strength of connectivity to other candidate cluster members. \\
\texttt{v\_l} & 2.78     & Longitudinal galactic velocity, $v_{l^*}$ [km\,s$^{-1}$]. \\
\texttt{v\_b} & -2.87     & Latitudinal galactic velocity [km\,s$^{-1}$]. \\
\texttt{x\_pc} & -8068.5  & Galactocentric $X$ position coordinate [pc]. \\
\texttt{y\_pc} & 256.0     & Galactocentric $Y$ position coordinate [pc]. \\
\texttt{z\_pc} & 86.3      & Galactocentric $Z$ position coordinate [pc]. \\
\texttt{(BP-RP)0} & -0.115 &  Gaia $G_\mathrm{BP}$-$G_\mathrm{RP}$   color, minus $E$($G_\mathrm{BP}$-$G_\mathrm{RP}$). \\
\texttt{(M\_G)0} & 0.442 & Absolute $G$-band magnitude, corrected for extinction. \\
\texttt{cluster} & CH-2 & RSG-5 or CH-2. \\
\texttt{Prot\_Adopted} & NaN & Adopted rotation period if available, else NaN [days]. \\
\texttt{Prot\_TESS} & NaN & TESS rotation period if available, else NaN [days]. \\
\texttt{Prot\_ZTF}  & NaN & ZTF rotation period if available, else NaN [days]. \\
\texttt{Prot\_Confused} & NaN & Boolean flag; true when stars are photometrically blended. \\
\enddata
\tablecomments{Table~3 is published in its entirety in a machine-readable
format.  One entry is shown for guidance regarding form and content.
}
\vspace{-0.5cm}
\end{deluxetable}

\begin{deluxetable*}{lllrrrrrrr}
	\tablecaption{ Priors and posteriors for the transit models with local
  polynomials removed.}
	\label{tab:koifull}
	\tabletypesize{\scriptsize}
	%
	\tablenum{4}
	\tablehead{
		\colhead{Param.} & 
		\colhead{Unit} &
		\colhead{Prior} & 
		\colhead{Median} & 
		\colhead{Mean} & 
		\colhead{Std{.} Dev.} &
		\colhead{3\% HDI} &
		\colhead{97\% HDI} &
		\colhead{ESS} &
		\colhead{$\hat{R}-1$}
	}
	\startdata
\hline
\multicolumn{10}{c}{\emph{Kepler-1643}} \\
\hline
$P$ & d & $\mathcal{N}(5.34264; 0.01000)$ & 5.3426257 & 5.3426258 & 0.0000101 & 5.3426071 & 5.3426454 & 7884 & 1.1e-03 \\
$t_0^{(1)}$ & d & $\mathcal{N}(134.38; 0.02)$ & 134.3820 & 134.3820 & 0.0011 & 134.3799 & 134.3841 & 7390 & 3.7e-04 \\
$\log R_{\rm p}/R_\star$ & -- & $\mathcal{U}(-6.215; 0.000)$ & -3.688 & -3.689 & 0.021 & -3.728 & -3.653 & 4449 & -7.8e-05 \\
$b$ & -- & $\mathcal{U}(0; 1+R_{\mathrm{p}}/R_\star)$ & 0.583 & 0.578 & 0.051 & 0.485 & 0.673 & 4705 & 1.9e-04 \\
$u_1$ & -- & \citet{exoplanet:kipping13} & 0.26 & 0.29 & 0.21 & 0.00 & 0.68 & 5324 & 7.9e-04 \\
$u_2$ & -- & \citet{exoplanet:kipping13} & 0.32 & 0.31 & 0.32 & -0.26 & 0.88 & 4908 & 8.4e-04 \\
$R_\star$ & $R_\odot$ & $\mathcal{N}(0.855; 0.044)$ & 0.851 & 0.851 & 0.045 & 0.766 & 0.933 & 7473 & 7.2e-04 \\
$\log g$ & cgs & $\mathcal{N}(4.502; 0.035)$ & 4.507 & 4.507 & 0.035 & 4.442 & 4.576 & 6530 & -1.4e-04 \\
$\log \sigma_f$ & -- & $\mathcal{N}(\log\langle \sigma_f \rangle; 2.000)$ & -8.520 & -8.520 & 0.019 & -8.556 & -8.486 & 7966 & 2.1e-04 \\
$\langle f \rangle$ & -- & $\mathcal{N}(1.000; 0.100)$ & 1.000 & 1.000 & 0.000 & 1.000 & 1.000 & 7488 & 3.2e-04 \\
$R_{\rm p}/R_\star$ & -- & -- & 0.025 & 0.025 & 0.001 & 0.024 & 0.026 & 4449 & -7.8e-05 \\
$\rho_\star$ & g$\ $cm$^{-3}$ & -- & 1.94 & 1.95 & 0.19 & 1.60 & 2.31 & 6081 & 9.4e-05 \\
$R_{\rm p}$ & $R_{\mathrm{Jup}}$ & -- & 0.207 & 0.207 & 0.012 & 0.184 & 0.231 & 6326 & 2.5e-04 \\
$R_{\rm p}$ & $R_{\mathrm{Earth}}$ & -- & 2.32 & 2.32 & 0.13 & 2.06 & 2.59 & 6326 & 2.5e-04 \\
$a/R_\star$ & -- & -- & 14.31 & 14.32 & 0.47 & 13.49 & 15.23 & 6081 & 8.2e-05 \\
$\cos i$ & -- & -- & 0.041 & 0.040 & 0.005 & 0.032 & 0.049 & 4929 & 2.4e-04 \\
$T_{14}$ & hr & -- & 2.41 & 2.41 & 0.06 & 2.30 & 2.53 & 4774 & 5.3e-04 \\
$T_{13}$ & hr & -- & 2.23 & 2.23 & 0.07 & 2.11 & 2.36 & 4561 & 6.2e-04 \\
\hline
\multicolumn{10}{c}{\emph{KOI-7368}} \\
\hline
$P$ & d & $\mathcal{N}(6.84294; 0.01000)$ & 6.8430344 & 6.8430341 & 0.0000125 & 6.8430107 & 6.8430574 & 10045 & 6.5e-05 \\
$t_0^{(1)}$ & d & $\mathcal{N}(137.06; 0.02)$ & 137.0463 & 137.0463 & 0.0014 & 137.0437 & 137.0489 & 10303 & 9.2e-05 \\
$\log R_{\rm p}/R_\star$ & -- & $\mathcal{U}(-4.605; 0.000)$ & -3.760 & -3.763 & 0.031 & -3.819 & -3.708 & 4043 & 6.3e-04 \\
$b$ & -- & $\mathcal{U}(0; 1+R_{\mathrm{p}}/R_\star)$ & 0.508 & 0.500 & 0.064 & 0.380 & 0.612 & 4434 & 3.5e-04 \\
$u_1$ & -- & \citet{exoplanet:kipping13} & 0.98 & 0.95 & 0.27 & 0.43 & 1.42 & 5809 & -5.6e-05 \\
$u_2$ & -- & \citet{exoplanet:kipping13} & -0.19 & -0.16 & 0.31 & -0.66 & 0.42 & 4387 & 2.6e-04 \\
$R_\star$ & $R_\odot$ & $\mathcal{N}(0.876; 0.035)$ & 0.874 & 0.874 & 0.036 & 0.804 & 0.938 & 9902 & 7.3e-04 \\
$\log g$ & cgs & $\mathcal{N}(4.499; 0.030)$ & 4.503 & 4.502 & 0.030 & 4.445 & 4.557 & 7527 & 2.7e-05 \\
$\log \sigma_f$ & -- & $\mathcal{N}(\log\langle \sigma_f \rangle; 2.000)$ & -8.314 & -8.314 & 0.012 & -8.337 & -8.292 & 10636 & 1.3e-03 \\
$\langle f \rangle$ & -- & $\mathcal{N}(1.000; 0.100)$ & 1.000 & 1.000 & 0.000 & 1.000 & 1.000 & 9742 & -2.9e-04 \\
$R_{\rm p}/R_\star$ & -- & -- & 0.023 & 0.023 & 0.001 & 0.022 & 0.025 & 4043 & 6.3e-04 \\
$\rho_\star$ & g$\ $cm$^{-3}$ & -- & 1.87 & 1.88 & 0.15 & 1.59 & 2.16 & 6829 & 3.4e-04 \\
$R_{\rm p}$ & $R_{\mathrm{Jup}}$ & -- & 0.198 & 0.198 & 0.011 & 0.177 & 0.218 & 5676 & 2.8e-04 \\
$R_{\rm p}$ & $R_{\mathrm{Earth}}$ & -- & 2.22 & 2.22 & 0.12 & 1.98 & 2.44 & 5676 & 2.8e-04 \\
$a/R_\star$ & -- & -- & 16.67 & 16.68 & 0.45 & 15.86 & 17.54 & 6829 & 3.3e-04 \\
$\cos i$ & -- & -- & 0.030 & 0.030 & 0.004 & 0.022 & 0.038 & 4518 & 5.4e-04 \\
$T_{14}$ & hr & -- & 2.79 & 2.79 & 0.07 & 2.65 & 2.93 & 4845 & 5.0e-04 \\
$T_{13}$ & hr & -- & 2.62 & 2.62 & 0.08 & 2.47 & 2.78 & 4575 & 3.1e-04 \\
\hline
\multicolumn{10}{c}{\emph{KOI-7913}} \\
\hline
$P$ & d & $\mathcal{N}(24.27838; 0.01000)$ & 24.278553 & 24.278571 & 0.000263 & 24.278112 & 24.279085 & 4413 & 1.5e-03 \\
$t_0^{(1)}$ & d & $\mathcal{N}(154.51; 0.05)$ & 154.5121 & 154.5124 & 0.0063 & 154.4998 & 154.5237 & 5612 & 6.0e-04 \\
$\log R_{\rm p}/R_\star$ & -- & $\mathcal{U}(-5.298; 0.000)$ & -3.599 & -3.602 & 0.046 & -3.689 & -3.519 & 4290 & 5.6e-04 \\
$b$ & -- & $\mathcal{U}(0; 1+R_{\mathrm{p}}/R_\star)$ & 0.312 & 0.298 & 0.153 & 0.005 & 0.523 & 2373 & 1.8e-03 \\
$u_1$ & -- & \citet{exoplanet:kipping13} & 0.27 & 0.34 & 0.28 & 0.00 & 0.86 & 4491 & -6.1e-05 \\
$u_2$ & -- & \citet{exoplanet:kipping13} & 0.21 & 0.23 & 0.32 & -0.31 & 0.86 & 5935 & 7.0e-04 \\
$R_\star$ & $R_\odot$ & $\mathcal{N}(0.790; 0.049)$ & 0.788 & 0.788 & 0.049 & 0.699 & 0.881 & 6847 & 2.8e-04 \\
$\log g$ & cgs & $\mathcal{N}(4.523; 0.043)$ & 4.526 & 4.527 & 0.042 & 4.450 & 4.606 & 5714 & 6.6e-04 \\
$\log \sigma_f$ & -- & $\mathcal{N}(\log\langle \sigma_f \rangle; 2.000)$ & -7.197 & -7.197 & 0.019 & -7.230 & -7.161 & 6976 & 1.4e-04 \\
$\langle f \rangle$ & -- & $\mathcal{N}(1.000; 0.100)$ & 1.000 & 1.000 & 0.000 & 1.000 & 1.000 & 6998 & 2.8e-04 \\
$R_{\rm p}/R_\star$ & -- & -- & 0.027 & 0.027 & 0.001 & 0.025 & 0.030 & 4290 & 5.6e-04 \\
$\rho_\star$ & g$\ $cm$^{-3}$ & -- & 2.20 & 2.21 & 0.25 & 1.78 & 2.70 & 5357 & 5.6e-04 \\
$R_{\rm p}$ & $R_{\mathrm{Jup}}$ & -- & 0.209 & 0.209 & 0.016 & 0.179 & 0.238 & 4882 & 1.3e-03 \\
$R_{\rm p}$ & $R_{\mathrm{Earth}}$ & -- & 2.34 & 2.34 & 0.18 & 2.01 & 2.67 & 4882 & 1.3e-03 \\
$a/R_\star$ & -- & -- & 40.92 & 40.95 & 1.54 & 38.14 & 43.84 & 5357 & 6.6e-04 \\
$\cos i$ & -- & -- & 0.008 & 0.007 & 0.004 & 0.000 & 0.013 & 2344 & 1.9e-03 \\
$T_{14}$ & hr & -- & 4.39 & 4.40 & 0.21 & 3.98 & 4.76 & 3952 & 5.6e-04 \\
$T_{13}$ & hr & -- & 4.13 & 4.13 & 0.22 & 3.72 & 4.55 & 3632 & 7.6e-04 \\
	\enddata
	\tablecomments{
		ESS refers to the number of effective samples.
		$\hat{R}$ is the Gelman-Rubin convergence diagnostic.
		Logarithms in this table are base-$e$.
		$\mathcal{U}$ denotes a uniform distribution,
		and $\mathcal{N}$ a normal distribution.
		\added{ Posterior values quoted in the text are means and standard
		deviations for symmetric distributions,
		and are otherwise medians bracketed by the
		upper and lower 84.1 and 15.9 percentile deviations.}
     (1) The ephemeris is in units of BJKD (BJDTDB-2454833).
	}
	\vspace{-0.3cm}
\end{deluxetable*}

\clearpage
\listofchanges
\end{document}